\documentclass{jfm}
\usepackage{graphicx}
\usepackage{gensymb}
\usepackage{epstopdf, epsfig}

\usepackage{subfig, float, amsmath, amsfonts, amssymb, bm}
\usepackage{siunitx} 
\usepackage{setspace} 
\usepackage{xcolor}
\usepackage{multirow}
\usepackage{tikz}

\DeclareRobustCommand\full  {\tikz[baseline=-0.6ex]\draw[very thick] (0,0)--(0.5,0);}
\DeclareRobustCommand\dotted{\tikz[baseline=-0.6ex]\draw[very thick,dotted] (0,0)--(0.54,0);}
\DeclareRobustCommand\dashed{\tikz[baseline=-0.6ex]\draw[very thick,dashed] (0,0)--(0.54,0);}
\DeclareRobustCommand\dashdot {\tikz[baseline=-0.6ex]\draw[very thick,dash dot] (0,0)--(0.5,0);}

\usetikzlibrary{shapes}
\newcommand{\markerone}{\raisebox{0.5pt}{\tikz{\node[draw,scale=0.4,thick,circle,fill=none](){};}}}

\newcommand{\blue}[1]{{\color{blue}#1}} 
\newcommand{\red}[1]{{\color{red}#1}} 
\newcommand{\orange}[1]{{\color{orange}#1}} 

\shorttitle{Effects of Phase Difference between Disturbances on Boundary Layer Transition}
\shortauthor{M. Kim, S. Kim, J. Lim, R.-S. Lin, S. Jee, and D. Park}

\title{Effects of Phase Difference between Instability Modes on Boundary Layer Transition}

\author{Minwoo Kim\aff{1}\footnote{M. Kim and S. Kim contributed equally to this work.}, Seungtae Kim\aff{1}$\dagger$, Jiseop Lim\aff{1}, Ray-Sing Lin\aff{1}, Solkeun Jee\aff{1}\corresp{\email{sjee@gist.ac.kr}} \and Donghun Park \aff{2}}

\affiliation{\aff{1}School of Mechanical Engineering, Gwangju Institute of Science and Technology, Gwangju 61005, Korea
\aff{2}Department of Aerospace Engineering, Pusan National University, Busan 46241, Korea}

\begin{document}

\maketitle

\begin{abstract}
Phase effect on the modal interaction of flow instabilities is investigated for laminar-to-turbulent transition in a flat-plate boundary layer flow. Primary and secondary instabilities are numerically studied with 2D Tollmien-Schlichting wave and subharmonic 3D oblique waves at various initial phase differences between these two instability modes. Three numerical methods are used for a systematic approach for the entire transition process, i.e., before the onset of transition well into fully turbulent flow. The Floquet analysis predicts the subharmonic resonance where a subharmonic mode locally resonates for a given basic flow composed of the steady laminar flow and the fundamental mode. Because the Floquet analysis is limited to the resonating subharmonic mode, nonlinear parabolized stability equations (PSE) simulation is conducted with various phase shifts of the subharmonic mode with respect to the given fundamental mode. PSE offers insights on the modal interaction affected by the phase difference up to the weakly nonlinear stage of transition. Large-eddy simulation (LES) is conducted for a complete transition to turbulent boundary layer because PSE becomes prohibitively expensive in the late nonlinear stage of transition. The modulation of the subharmonic resonance with the initial phase difference leads to a significant delay in the transition location up to $\Delta Re_{x, tr} \simeq 4\times 10^5$ as predicted by the current LES. Effects of the initial phase difference on the spatial evolution of the modal shape of the subharmonic mode are further investigated. The mechanism of the phase evolution is discussed, based on current numerical results and relevant literature data.
\end{abstract}

\section{Introduction}\label{sec:intro}
	
	Laminar-to-turbulent transition may occur in boundary layer flow after instability modes in the laminar region interact each other. Challenges in assessing the transition mechanism arise owing to the nonlinear (or energetic) interaction.  Various transition routes were suggested for wall-bounded flow by \citet{Morkovin1969} about five decades ago. After disturbances outside the boundary layer converts to primary instability in the boundary layer through the receptivity process, secondary instability normally occurs as the primary instability grows to sufficiently large amplitude and promotes modal interactions. The secondary instability (here, subharmonic resonance) refers to the amplification of a 3D wave catalyzed by a 2D Tollmien-Schlichting (TS) wave, i.e., the primary instability mode of the Orr-Sommerfeld flow field. The 3D wave has one-half the frequency of the fundamental TS wave.

	Significant progresses are made in the understanding of secondary instability, which are well reviewed in \citet{Herbert1988, Kachanov1994, Saric2003, Schmid2007}. Subharmonic secondary instability has been investigated extensively, using theoretical approaches \citep{Craik1971, Herbert1984, Herbert1988, Wu2019}, experimental studies \citep{Kachanov1984, Corke1989, Borodulin2002, Wuerz2012}, and numerical investigation \citep{El-Hady1988, Nayfeh1990, Bertolotti1992, Joslin1993, Xu2017, Jee2018, Kim2019, Kim2020}. The current study focuses on the subharmonic resonance in the natural transition path for incompressible flow. Other transition routes including non-modal interaction leading to transient growth and bypass transition are well reviewed by \citet{Schmid2007} and \citet{Durbin2007}, respectively. 
	
	It is well recognized that the early stage of secondary instability involves the parametric resonance of the subharmonic mode with respect to the basic flow composed of steady slow-varying laminar flow and the 2D fundamental mode \citep{Herbert1984, Herbert1988, El-Hady1988, Nayfeh1990}. The parametric resonance of the subharmonic mode can be analyzed with the Floquet theory applied to the basic flow. Among several parameters of instability modes affecting the growth of the subharmonic mode, key parameters have been identified as the local Reynolds number of the basic flow, the amplitude of the fundamental mode, and the spanwise wavenumber of the subharmonic mode \citep{Herbert1988}. The Floquet analysis for subharmonic resonance \citep{Herbert1984, Herbert1987, Nayfeh1990} has been validated for well-controlled experiments of \citet{Klebanoff1962, Kachanov1984}.
	
	Despite of the vast literature on the subharmonic resonance, the complete understanding of the nonlinear interaction between the fundamental and the subharmonic modes has not been achieved.  Particularly, the nonlinear interaction influenced by the phase difference between the two waves has not gained enough attention in the research community. Recently, \citet{Park2019} reproduced the phase-dependent subharmonic resonance observed in the experiment of \citet{Borodulin2002} using the nonlinear parabolized stability equations (PSE) analysis. Yet, the effect of the phase difference on the subharmonic resonance has not been fully identified. Note that previous studies \citep{Borodulin2002, Wuerz2012, Park2019} were still confined in the pre-turbulence region due to experimental \citep{Borodulin2002, Wuerz2012} and numerical \citep{Park2019} constraints. 
	
	The goal of the current study is to improve the understanding of the phase effect on boundary layer transition, covering a wide transition range from the early stage of primary and secondary instabilities to turbulent flow. To achieve such a comprehensive investigation, three numerical methods are judiciously incorporated: the Floquet analysis, PSE, and large-eddy simulation (LES). The Floquet analysis provides the resonating subharmonic mode (secondary instability) with respect to the fundamental mode (primary instability). An in-house code is developed for the current Floquet analysis based on previous studies \citep{El-Hady1988, Nayfeh1990}. Since the Floquet analysis is limited to the resonant subharmonic mode (so resonant phase difference), PSE is chosen as a higher-fidelity stability analysis for non-resonant phase differences. Because PSE can handle the nonlinearity and non-parallelism of the disturbance equations, the instability evolution affected by the phase can be investigated in the nonlinear transition region. A well-validated nonlinear PSE code by \citet{Park2013, Park2016, Park2019} is used in the current study. Although PSE is an effective method for stability analysis in a pre-turbulence region, it is computationally prohibitive in turbulent flow. The authors have developed an LES method coupled with the stability analysis \citep{Kim2019, Kim2020, Lim2021} for a cost-effective and high-fidelity simulation of transitional boundary layer. This LES method is used for a complete turbulent transition here, and the transition location controlled by the initial phase difference is quantified. It should be noted that the current LES approach \citep{Kim2019, Kim2020} provides the direct numerical simulation (DNS) fidelity in the laminar region where disturbances are deterministic.
	
	In Section \ref{sec:methods}, three methods, the Floquet analysis, PSE and LES are described. The validation of the current Floquet analysis is discussed in Section \ref{sec:results-validation}. The effect of the phase difference on the subharmonic resonance is thoroughly investigated using the PSE and LES in Section \ref{sec:pse-only} and \ref{sec:les}, respectively. The mechanism of the phase synchronisation from the anti-resonant condition is further discussed in Section \ref{sec:evolution}. A summary and conclusions are given in Section \ref{sec:conclusions}.

\section{Methods \label{sec:methods}}

	Three numerical methods are used in this study. Parametric resonance of the subharmonic oblique wave is investigated using the Floquet analysis. PSE computations are conducted to study the effect of the phase difference on the parametric resonance. LES is carried out to simulate complete transition to turbulent boundary layer. The Floquet analysis, PSE and LES are briefly discussed in Sections \ref{sec:Floquet}, \ref{sec:PSE} and \ref{sec:LES}, respectively. 
	
	In this study, a total variable $\check{\psi}$ is decomposed to the undisturbed part $\Psi$ and the disturbance $\psi$.
	\begin{equation}
	\check{\psi} = \Psi + \psi
	\label{eq:psi}
	\end{equation}
	The Cartesian coordinate system is used with the streamwise $x$, wall-normal $y$, and spanwise $z$ direction, along with the corresponding velocity components $u$, $v$, and $w$. Dimensionless variables are obtained with the length scale $\tilde{\delta}_r(R=R_o)$, the freestream velocity $\tilde{U}_\infty$, and the dynamic pressure $\tilde{\rho}\tilde{U}^2_\infty$, where the local length variable is $\tilde{\delta}_r=\sqrt{\tilde{x}_r \tilde{\nu} / \tilde{U}_\infty}$, $\tilde{x}_r $ is the distance from the leading edge of a flat plate, $\tilde{\nu}$ is kinematic viscosity, the local Reynolds number is $R=\tilde{U}_\infty \tilde{\delta}_r/\tilde{\nu}=\sqrt{\tilde{U}_\infty \tilde{x}_r/\tilde{\nu}}=\sqrt{Re_x}$, the reference $R$ is $R_o = 400$, and $\tilde{\rho}$ is the fluid density. The tilde denotes a dimensional variable. 

\subsection{Floquet analysis \label{sec:Floquet}}
	
	The Floquet analysis is based on the parametric formulation which describes secondary instability for a given primary instability \citep{Herbert1984, Herbert1987, Nayfeh1990, El-Hady1988}. The current Floquet analysis, briefly described here, adopts the approach of \citet{Nayfeh1990, El-Hady1988} instead of the stream-function approach of \citet{Herbert1984, Herbert1987}.

	The governing equations of a disturbance ($u, v, w, p$) for a undisturbed basic flow ($U, V$) are written as	
	\begin{subeqnarray}
	 \frac{\partial u}{\partial x} + \frac{\partial v}{\partial y} + \frac{\partial w}{\partial z} &=&0  \\
	 \frac{\partial u}{\partial t} + U\frac{\partial u}{\partial x} + v\frac{\partial U}{\partial y} + \frac{\partial p}{\partial x} - \frac{1}{R_o}\nabla^2 u 
	 + \left[u\frac{\partial U}{\partial x} + V\frac{\partial u}{\partial y}  \right] + \mathbb{N}_u &=&0 \\
	 \frac{\partial v}{\partial t}+U\frac{\partial v}{\partial x}+\frac{\partial p}{\partial y} - \frac{1}{R_o}\nabla^2 v 
	 + \left[ u\frac{\partial V}{\partial x} + V\frac{\partial v}{\partial y} + v\frac{\partial V}{\partial y} \right] + \mathbb{N}_v &=& 0 \\
	 \frac{\partial w}{\partial t} + U\frac{\partial w}{\partial x} + \frac{\partial p}{\partial z} - \frac{1}{R_o}\nabla^2 w 
	 + \left[ V\frac{\partial w}{\partial y} \right] + \mathbb{N}_w &=&0 
	 \label{eq:disturb}
	\end{subeqnarray}
where the square-bracket terms represent the non-parallelism of basic flow. Nonlinear terms of disturbance $\mathbb{N}$'s are negligible here.
	
	The current instability analysis involves two steps: primary and secondary instability. The primary instability analysis provides a 2D fundamental TS wave for a given basic laminar flow, whereas the secondary instability analysis (here Floquet analysis) yields a 3D subharmonic wave for a given basic flow in which the 2D wave is additionally included. The current analysis is summarized in table \ref{tab:basic_dist} with a brief description below. 

	\begin{table}
	\centering
	\small{
	\begin{tabular}{c c c c}
	Analysis type					& Basic flow		 		& Disturbance 						& Equation for  						\\
								& (given)				& (unknown)							& analysis							\\
	\hline
	\:{ } Primary instability \:{ } 	& Laminar solution 		& \:{ } 2D fundamental TS wave \:{ } 	& \multirow{2}{*}{Eq. \ref{eq:Pri}} 	\\
	analysis						& Eq. \ref{eq:2D_basic}	& Eq. \ref{eq:2D_dist} 				&									\\
	\hline
	\:{ } Secondary instability \:{ }	& Laminar + 2D TS wave 	& \:{ } 3D subharmonic wave \:{ }		&  \multirow{2}{*}{Eq. \ref{eq:Sec}}	\\
	analysis (Floquet)				& Eq. \ref{eq:3D_basic} 	& Eq. \ref{eq:3D_dist}				&									\\
	\end{tabular}
	}
	\caption{Basic flow and disturbance for each analysis of primary and secondary instability.}
	\label{tab:basic_dist}
	\end{table}

	For the primary instability of boundary layer, we consider the basic flow  
	\begin{equation}
	\left\{ U, V \right\} = \left\{ U_L(y), 0 \right\} 
	\label{eq:2D_basic}
	\end{equation}
	where $U_L$ is the laminar solution without any disturbance (here the Blasius solution).  Then, a fundamental planar TS wave can be written as 
	\begin{equation}
	\left\{ u,v,p \right\} = \left\{ \zeta_1(y), \zeta_3(y), \zeta_4(y) \right\} \exp \left[ i \left( \alpha x -\omega t \right) \right] + \text{c.c.} 
	\label{eq:2D_dist}
 	\end{equation}
	where the notation c.c. indicates the complex conjugate. The functions $\zeta_1(y)$, $\zeta_3(y)$ and $\zeta_4(y)$ are the mode shape of the fundamental TS components $u$, $v$ and $p$, respectively. For a spatially evolving disturbance, the complex wavenumber $\alpha$ and the real angular frequency $\omega$ are used.   
	
	To obtain the subharmonic oblique wave in the Floquet analysis, the basic flow (Eq. \ref{eq:3D_basic}) includes both the laminar flow and the fundamental 2D TS wave.
	\begin{subeqnarray}
		U &=& U_L(y) + A \left[ \zeta_1 (y) \exp(i\theta) + \zeta^*_1  (y) \exp(-i\theta) \right] \\
		V &=& A \left[ \zeta_3  (y) \exp(i\theta) + \zeta^*_3  (y) \exp(-i\theta) \right] 
	\label{eq:3D_basic}
	\end{subeqnarray}
	where $A$ is the rms amplitude of the fundamental TS wave, $\zeta^*$ is the complex conjugate of the function $\zeta$, and $\theta(x,t)=\Re(\alpha)x-\omega t$.  The notations $\Re$ and $\Im$ indicate the real and the imaginary part of a complex variable, respectively. The Floquet theory suggests that the approximate solution of Eq. \ref{eq:disturb} for the given basic flow of Eq. \ref{eq:3D_basic} can be written as
	\begin{subeqnarray}
		\arraycolsep=0pt
		\left\{ \begin{array}{c}
		u_{1/2} \\
		v_{1/2} \\
		p_{1/2} \\ \end{array} \right\} &=&
		\left[ 
		\arraycolsep=0pt
		\left\{ \begin{array}{c}
		\eta_1(y) \\
		\eta_3(y) \\ 
		\eta_4(y) \\ \end{array} \right\} \exp(i\theta/2) + 
		\arraycolsep=0pt
		\left\{ \begin{array}{c}
		\eta_7(y) \\
		\eta_9(y) \\ 
		\eta_{10}(y) \\ \end{array} \right\} \exp(-i\theta/2)
		\right] \cos(\beta z) \exp(\gamma x)  \\
		w_{1/2} &=& \left[ \eta_5(y) \exp(i\theta/2) + \eta_{11}(y) \exp(-i\theta/2) \right] \sin(\beta z) \exp(\gamma x) 
	\label{eq:3D_dist}
	\end{subeqnarray}	
	where the subharmonic wavenumber in the streamwise direction is $\alpha_{1/2}=\Re(\alpha)/2$, the subharmonic frequency is $\omega_{1/2}=\omega/2$ (so, $\theta/2=\alpha_{1/2}x-\omega_{1/2}t$), the spanwise wavenumber of the subharmonic mode is $\beta$, and the eigenvalue $\gamma$ is real here. Further details are documented in Appendix~\ref{appendix}, including the exact equation for each analysis, boundary conditions, and the computational method for the eigenvalue problems.
	
\subsection{Parabolized stability equations (PSE) \label{sec:PSE}}

	The nonlinear parabolized stability equations (PSE) is an efficient method for a weakly nonlinear region where, for parametric resonance occurs here. The amplitude of the subharmonic mode remains small so that the back influence of the subharmonic mode on the fundamental mode is negligible. A formal approach of PSE is to decompose the disturbance $\psi$ into a fast-varying oscillatory-wave part and a slow-varying shape function, using Fourier expansion, as written in Eq. \ref{eq:PSE}.
	\begin{equation}
	\psi(x,y,z,t) = \sum_{m=-N_m}^{N_m} \sum_{n=-N_n}^{N_n} \hat\psi_{(m,n)} (x, y) 
	\exp \left[i  \left\{ \int_{x_{0}}^{x} \alpha_{(m,n)}(s) d s + n \beta z - \frac{1}{2} m \omega t  \right\} \right]
	\label{eq:PSE}
	\end{equation}
	where the shape function $\hat\psi_{(m,n)}$ is a complex function, the streamwise wavenumber $\alpha_{(m,n)}$ is a complex number, the spanwise wavenumber $\beta$ is a real number. The subscript $m$ and $n$ indicates the temporal and spanwise modes, respectively. The wavenumber $\alpha_{(2,0)}$ of the fundamental TS wave corresponds to the wavenumber $\alpha$ in Eq. \ref{eq:2D_dist}. The subharmonic wavenumber $\alpha_{(1,1)}$ is associated to the notations $\alpha_{1/2}$ and $\gamma$ in Eq. \ref{eq:3D_dist}, i.e., $\alpha_{(1,1)}=\alpha_{1/2} - i \gamma$. The spatial growth rate of the subharmonic mode is $\Im(\alpha_{(1,1)})=-\Re(\gamma)$. 
	
	A set of partial differential equations for the shape functions with the unknown variable $\alpha_{(m,n)}$ is obtained for a given frequency (here the fundamental frequency $\omega$) and the spanwise wavenumber $\beta$. These equations are parabolized and numerically solved in the PSE code developed in \citet{Park2013, Park2016}. Nonlinear PSE is conducted with $N_m=6$ and $N_n=3$, keeping total 28 modes including the mean distortion $(0,0)$ mode, in the domain of $400 \leq R \leq 700$. The 4th-order central scheme and the 2nd-order backward scheme are used for the wall-normal and the streamwise direction, respectively. Uniform grids with 107 points are used in the streamwise direction. At least 80 points are placed in the boundary layer with a total 220 points in the wall-normal domain, extending to $200 \tilde{\delta}_r$. Further details of the PSE code and numerical approaches are documented in \citet{Park2013, Park2016, Kim2019, Park2019}.

	It should be noted that the PSE code is based on the compressible form of the disturbance equation. To approximate the incompressible boundary layer, the mean flow of very low freestream speed whose Mach number is 0.0269 is chosen (i.e., 9.16 m/s with the standard atmospheric condition at sea level). At this Mach number condition and the corresponding mean flow data, the density and temperature fluctuations behave as redundant variables in the PSE analysis. Although the compressible formulation is used and all terms are kept, the results can be regarded as nearly identical to those obtained from the incompressible formulation as they have been validated through many cases \citep{Bertolotti1992, Chang1993, Gao2011, Park2013}. Besides, the results from the Floquet theory are used as the inlet boundary conditions for PSE as well as LES. For the initial condition of PSE, the pressure and velocity disturbances are set as Floquet theory results, while the density and temperature disturbances are set as zero. As a consequence, there can be a small difference in comparison with the solution with the compressible formulation. However, this discrepancy is also almost negligible owing to the very low Mach number considered.

\subsection{Large-eddy simulation (LES) \label{sec:LES}}

	Since both the Floquet analysis and PSE are not efficient for incorporating higher instability in the late stage of the transition, large-eddy simulation (LES) is conducted to model the boundary-layer flow well into a fully turbulent state. The total variable $\check{\psi}$ is decomposed to the spatially filtered $\bar\psi$ and the filtered-out, residual part $\psi'$ for LES. 
	\begin{equation}
	\check{\psi} = \bar{\psi} + \psi'
	\end{equation}
Note that there is no distinction between $\check{\psi}$ and $\bar\psi$ in the computation of laminar flow because LES is essentially DNS there \citep{Kim2019, Kim2020}. The residual part $\psi'$ is a turbulence fluctuation, so $\psi'=0$ in the pre-transition region where the deterministic disturbance $\psi$ (see Eq.~\ref{eq:psi}) is well resolved with the current fine grid.

	The filtered incompressible dimensionless Navier-Stokes equations are written as Eq. \ref{eq:LES}. 
	\begin{subeqnarray}
	\frac{\partial \bar{u}_i}{\partial x_i} &=& 0 \\
	\frac{\partial \bar{u}_j}{\partial t} + \frac{\partial}{\partial x_i} \left( \bar{u}_i \bar{u}_j \right) &=&
	\frac{1}{R} \frac{\partial ^2 \bar{u}_j}{\partial x_i \partial x_i} - \frac{\partial \tau^{r}_{ij}}{\partial x_i} - \frac{\partial \bar{p}}{\partial x_j} \\
	\tau^r _{ij} &=& -2\nu _t \bar{S}_{ij} \\
	\bar{S}_{ij} &=& \frac{1}{2} \left( \frac{\partial \bar{u}_i}{\partial x_j} + \frac{\partial \bar{u}_j}{\partial x_i} \right)
	\label{eq:LES}
	\end{subeqnarray}
	where the residual stress tensor $\tau^r _{ij}$ is modeled to be linearly proportional to the resolved strain rate $\bar{S}_{ij}$. The turbulence viscosity $\nu_t$ is obtained with the wall-adapting local eddy-viscosity (WALE) model \citep[see][]{Nicoud1999}. The model coefficient is $C_w=0.5$ as suggested in \citet{Nicoud1999} and tested in the transitional boundary layer in \citet{Kim2019, Kim2020}. The sub-grid-scale turbulence model is judiciously chosen for wall-resolved LES in transitional boundary layer \citep{Kim2020} --- the model is properly activated only in the very late transition stage and the subsequent turbulent region in the current simulation.
	
\begin{figure}
	\centering
	\includegraphics[width=.8\textwidth]{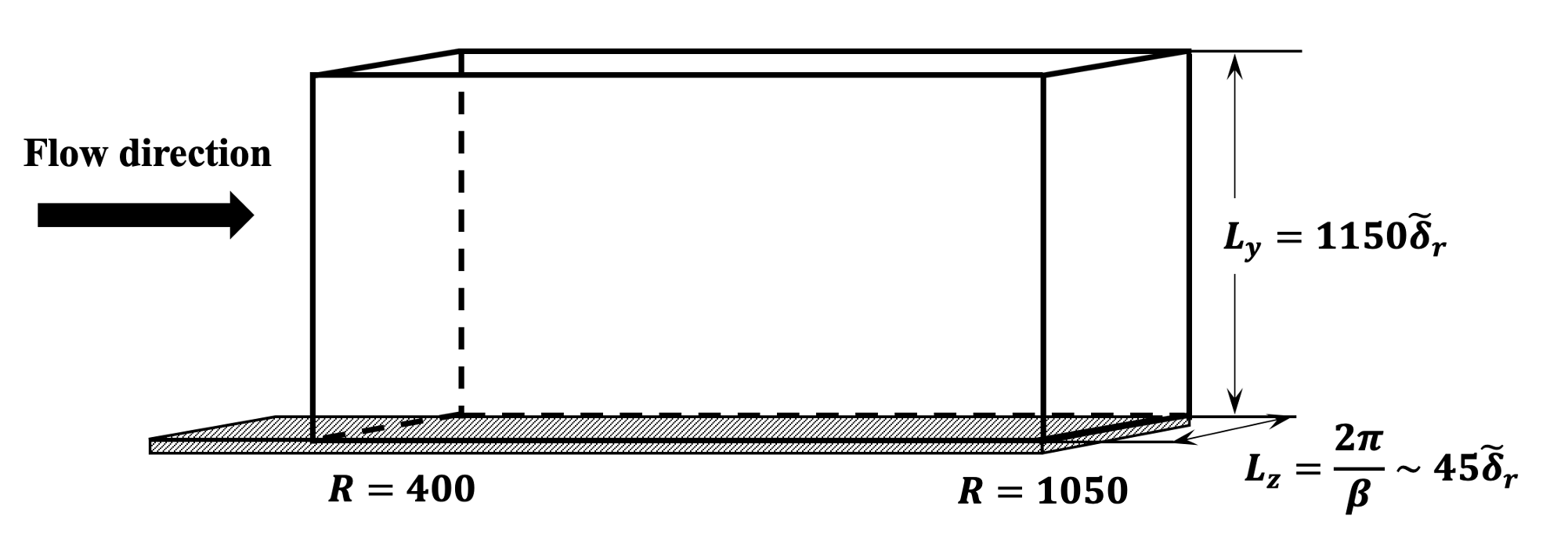}
	\caption{Schematic diagram of the current LES computational domain.}
	\label{fig:LES_domain}
\end{figure}

	The computational domain is depicted in figure \ref{fig:LES_domain}. At the LES inlet, the fundamental and subharmonic modes are assigned to the laminar solution. The convective outlet is applied at the exit boundary, and no-slip condition for the wall. The fine grid of \citet{Kim2019} is used here with the small enough time step $\Delta_t=\frac{1}{256}\frac{2\pi}{\omega}$ where $\omega$ is the angular frequency of the fundamental mode. Further details on the LES approach are referred to \citet{Kim2019, Kim2020}.
	
\subsection{Inlet conditions for PSE and LES \label{sec:inlet}}

Both PSE and LES computations require a disturbance at the inlet boundary, which is located at $R=400$. The 2D TS wave and the pair of the 3D oblique waves are obtained from the current Floquet analysis. Following the Floquet analysis of \citet{Herbert1987}, which has been validated for the experiment of \citet{Kachanov1984}, the angular frequency of the 2D wave is $\omega=0.0496$ and the spanwise wavenumber of the 3D wave is $\beta=0.132$ based on the current non-dimensionalization. These two parameters can be re-written as $F=\omega / R \times 10^6=124$ and $b=\beta / R \times 10^3=0.33$, respectively, which are also commonly used in the literature. The rms amplitudes of the 2D wave and the 3D wave are $A_{(2,0)}=4.0\times 10^{-3}$ and  $A_{(1,1)}=1.64\times 10^{-5}$, respectively, at the inlet, where the freestream velocity is used for the scaling. At the LES and PSE inlet boundary, 2D fundamental and 3D subharmonic modes are added to the laminar solution; namely a zero-pressure-gradient flat-plate flow at $R=400$.

\begin{figure}
	\centering
	\subfloat[Amplitude $|\zeta|/\max(|\zeta|)$]{ \includegraphics[width=.45\textwidth]{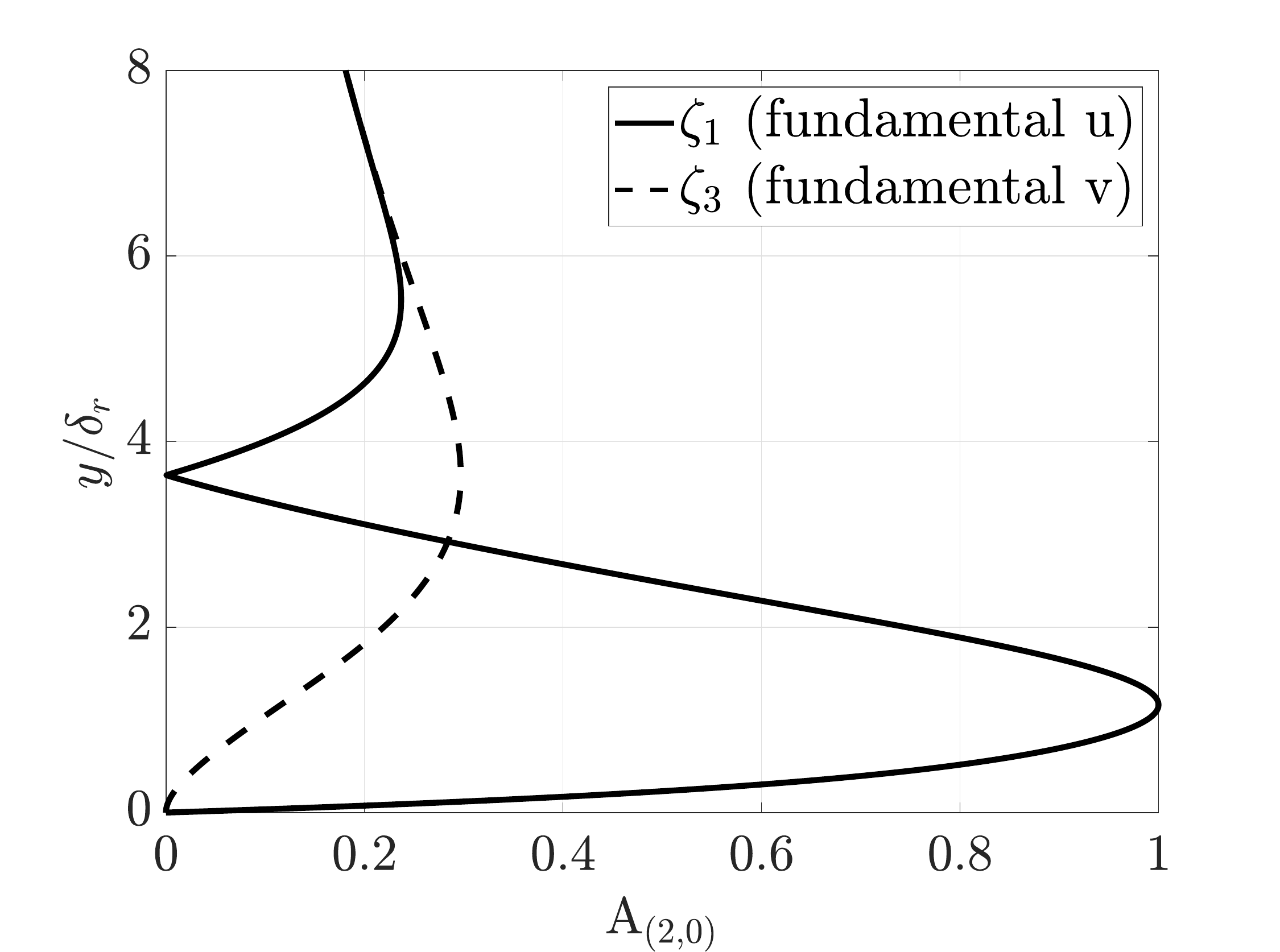}	}
	\subfloat[Phase $\phi_{(2,0)}=\arg(\zeta)$]{	\includegraphics[width=.45\textwidth]{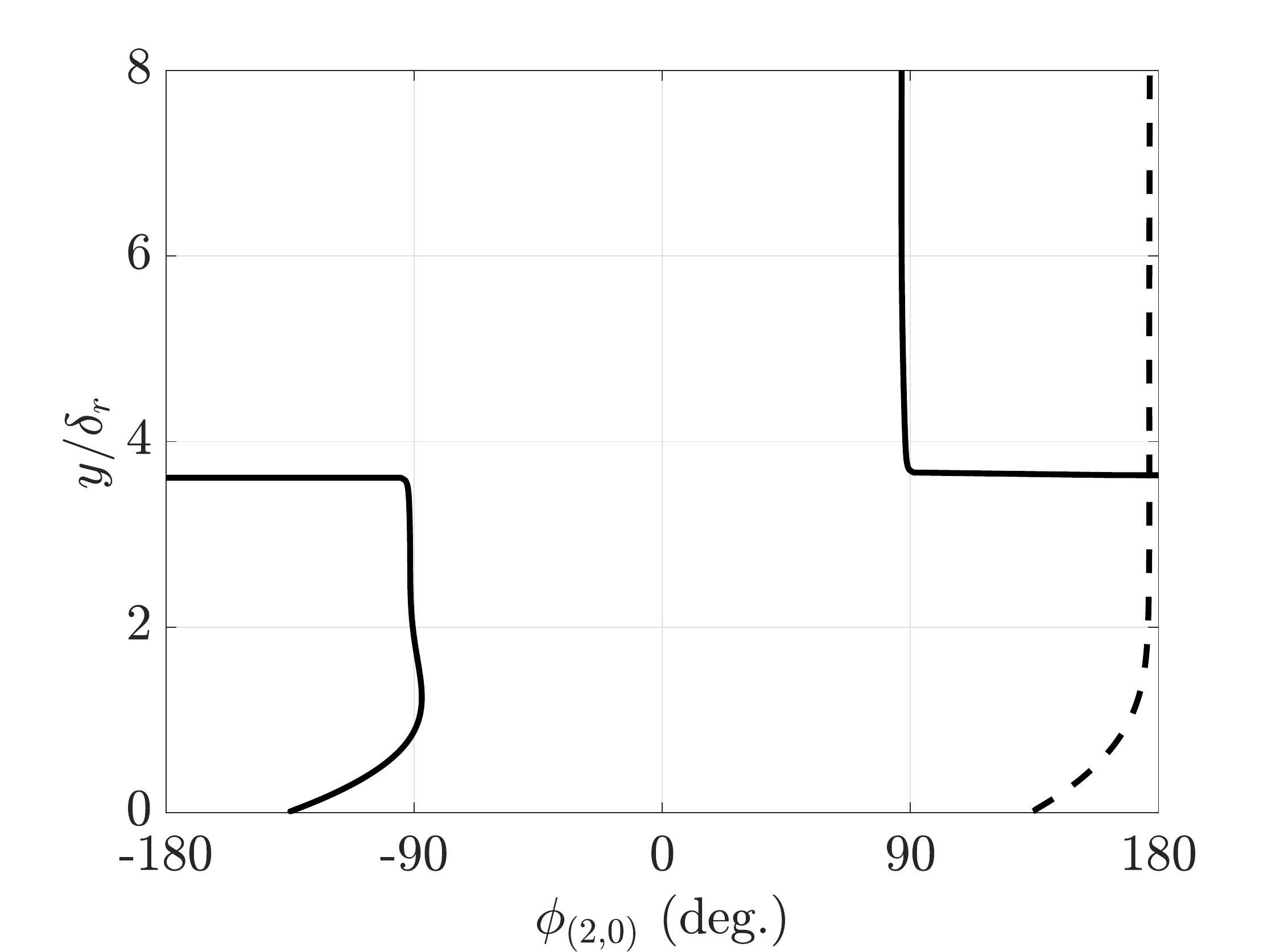}		}\\
	\subfloat[Amplitude $|\eta|/\max(|\eta|)$]{ \includegraphics[width=.45\textwidth]{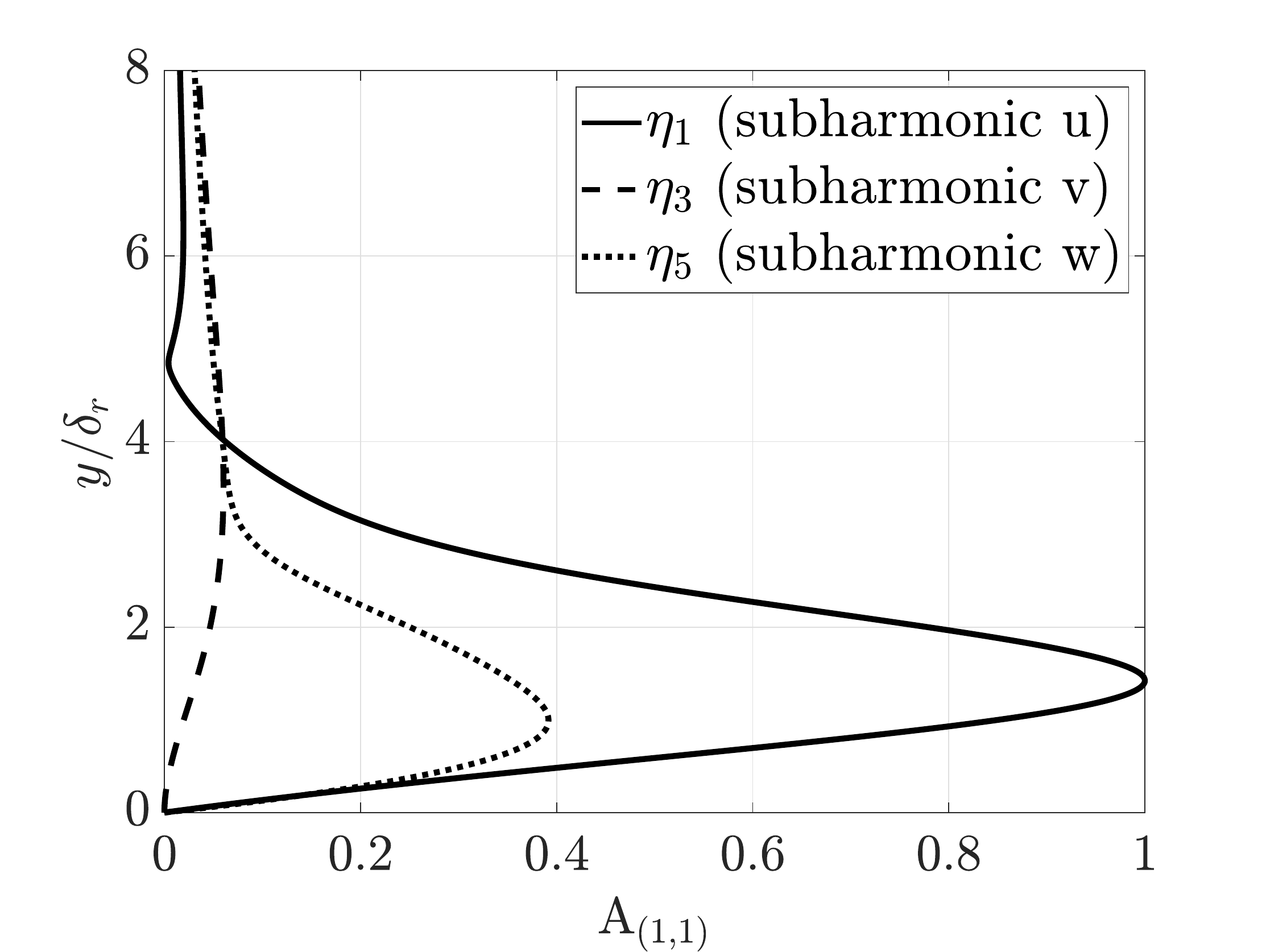}	}
	\subfloat[Phase $\phi_{(1,1)}=\arg(\eta)$]{	\includegraphics[width=.45\textwidth]{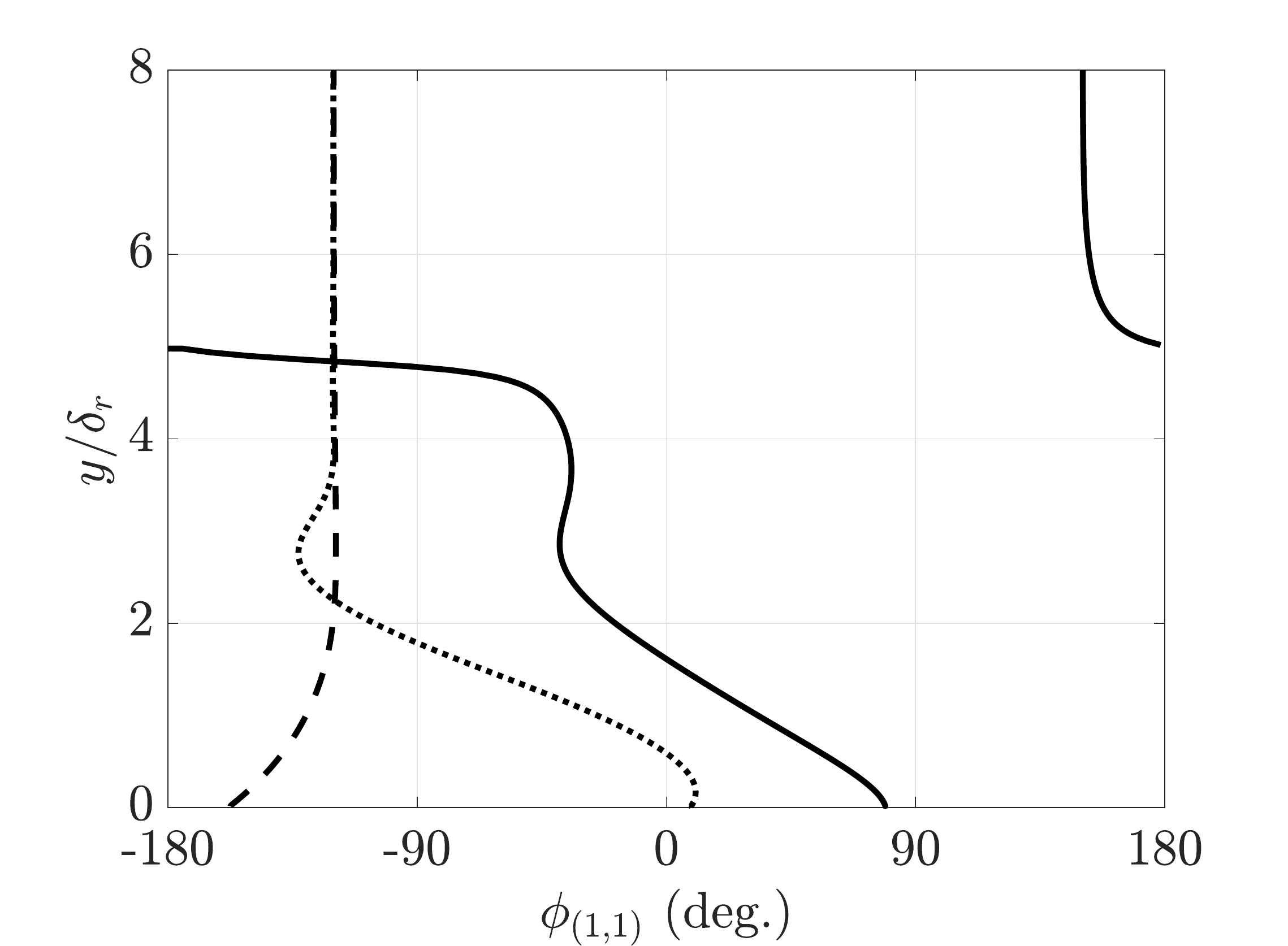}		}
	\caption{The fundamental and the subharmonic modes obtained from the current Floquet analysis at $R=400$.}
	\label{fig:inlet}
\end{figure}

The current Floquet analysis provides the mode shape of the fundamental (2D TS wave) $\zeta$ and the subharmonic (3D oblique wave) $\eta$ modes, as shown in figure~\ref{fig:inlet}. The amplitude of the each mode is scaled with the maximum value of each mode. The $u$ component dominates both the fundamental and subharmonic modes. The amplitude peak of the fundamental and subharmonic modes is located at about one quarter of the boundary thickness. The phase profile of the fundamental mode $\zeta_1$ is relatively a constant around the amplitude peak, whereas the phase of $\eta_1$ changes continuously.  

The phase difference $\Delta \phi$ between the fundamental and the subharmonic modes is defined as Eq.~\ref{eq:phase_difference}, following experimental studies of \citet{Borodulin2002, Wuerz2012}.
\begin{equation}
	\Delta \phi = \frac{1}{2} \phi_{(2,0)} - \phi _{(1,1)} \;\; \text{at} \;\; y_{(1,1),\max} 
	\label{eq:phase_difference}
\end{equation}
where $y_{(1,1), \max}$ is the location for the amplitude peak of the subharmonic mode which is $y_{(1,1), max}=1.43$ at the inlet. The initial phase difference between the fundamental and subharmonic modes is $\Delta \phi _{in}=130 \degree$ from the Floquet analysis. In the current PSE and LES computations, $\Delta \phi _{in}$ varies in the periodic range of 180 degrees in order to investigate the effect of the phase difference on secondary instability and eventually on the turbulent transition. The subharmonic phase is shifted with respect to the given fundamental mode. There is no distinction between the phase lead and lag results in the periodicity of 180 degrees for the phase difference.

\section{Results}\label{sec:results}

The effect of the modal phase on the growth of the secondary instability (here, the subharmonic mode) is investigated using three approaches, i.e., the Floquet analysis, PSE and LES, in this study. The baseline case is the subharmonic resonance in the zero-pressure-gradient boundary layer on a flat plate, which was experimentally studied by \citet{Kachanov1984}. The current Floquet analysis, validated against the experimental \citep{Kachanov1984} and the numerical data \citep{Herbert1987} in Section~\ref{sec:results-validation}, provides a resonating subharmonic mode for the given basic flow which consists of the laminar flow and the fundamental mode (2D TS wave).  Because of the nature of the eigenvalue problem explored in the Floquet analysis, other methods are required for less-resonating conditions affected by the modal phase. Here, PSE and LES are used. The phase effect is first discussed with PSE in Section~\ref{sec:pse-only} and then with LES in Section~\ref{sec:les}. A further investigation on the transition location delay affected by the phase and the resonant mechanism from an anti-resonant initial condition is included in Section~\ref{sec:evolution}.

\subsection{Validation of instability analysis \label{sec:results-validation}}

The current instability analysis consists of two parts, one for the primary instability (2D fundamental TS wave) and the other for the secondary instability (3D subharmonic oblique wave). The eigenvalue problem of each analysis is summarized in table~\ref{tab:basic_dist} along with the given basic flow. The Floquet analysis provides the most unstable mode for the subharmonic oblique wave, yielding the mode shape $\eta$ and the exponent $\gamma$ whose positive real value is the spatial growth rate of the subharmonic wave. 

\begin{figure}
	\centering
	\subfloat[Fundamental mode]{ \includegraphics[width=.4\textwidth]{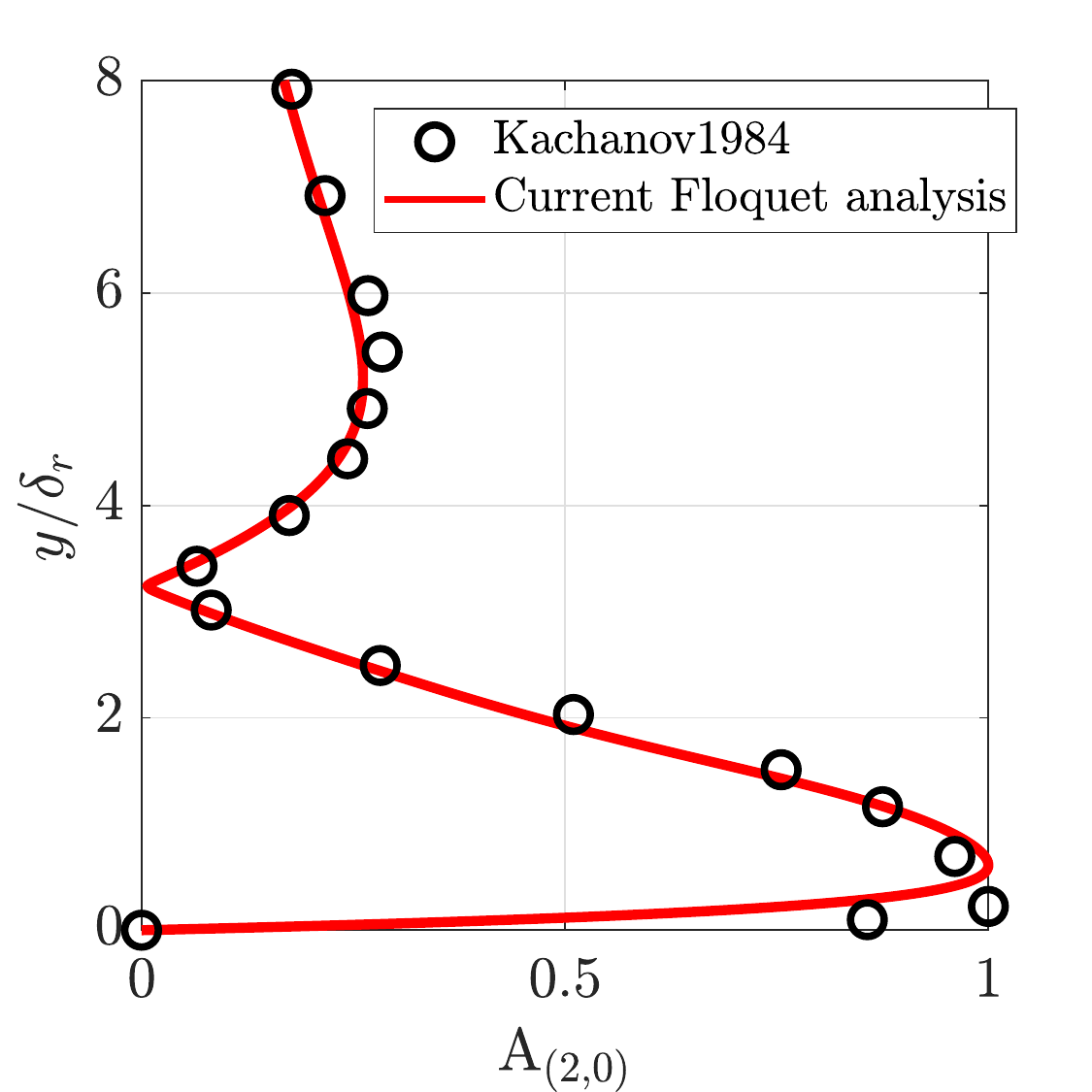}} 
	\subfloat[Subharmonic mode]{ \includegraphics[width=.4\textwidth]{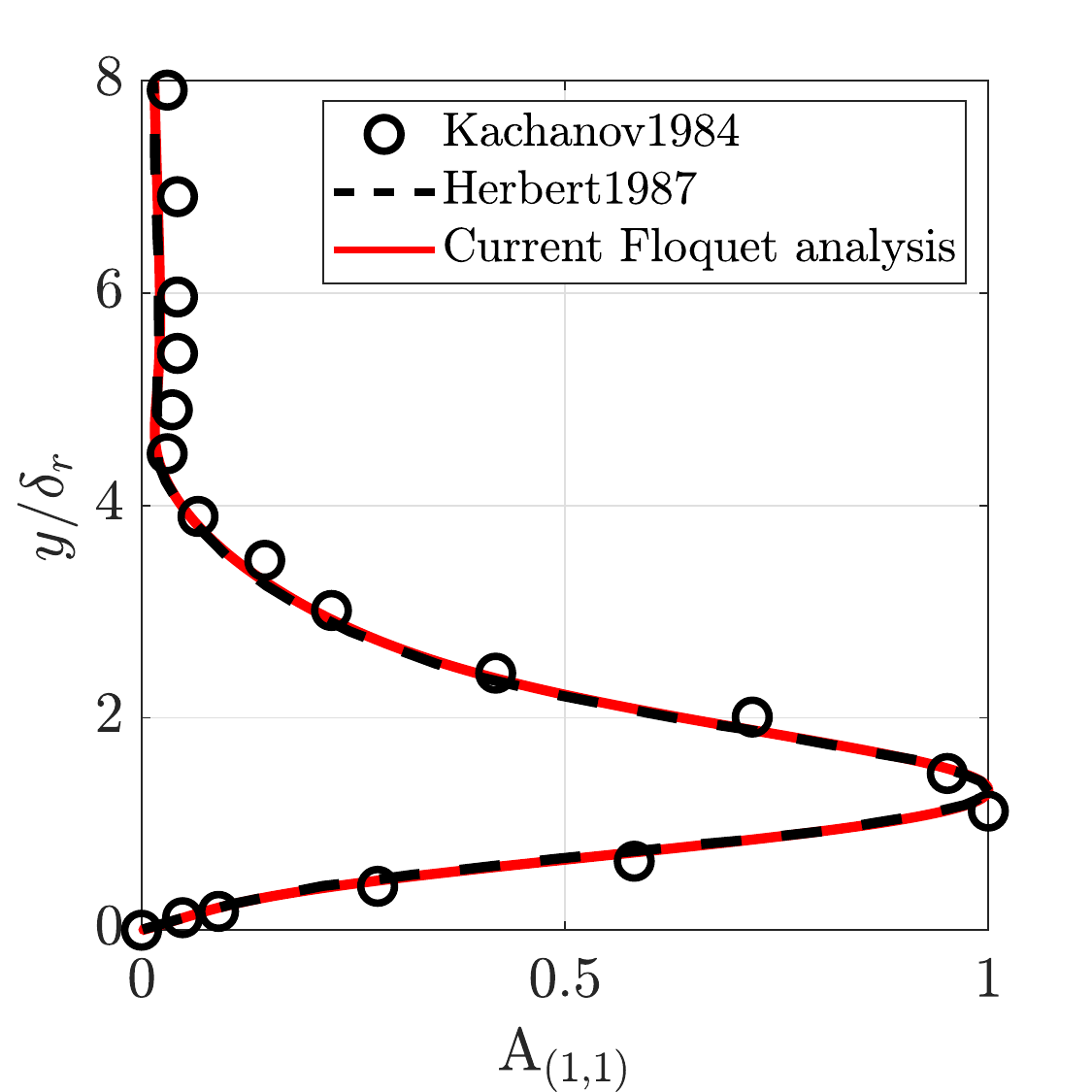} } \\
	\subfloat[Amplitude growth]{ \includegraphics[width=.6\textwidth]{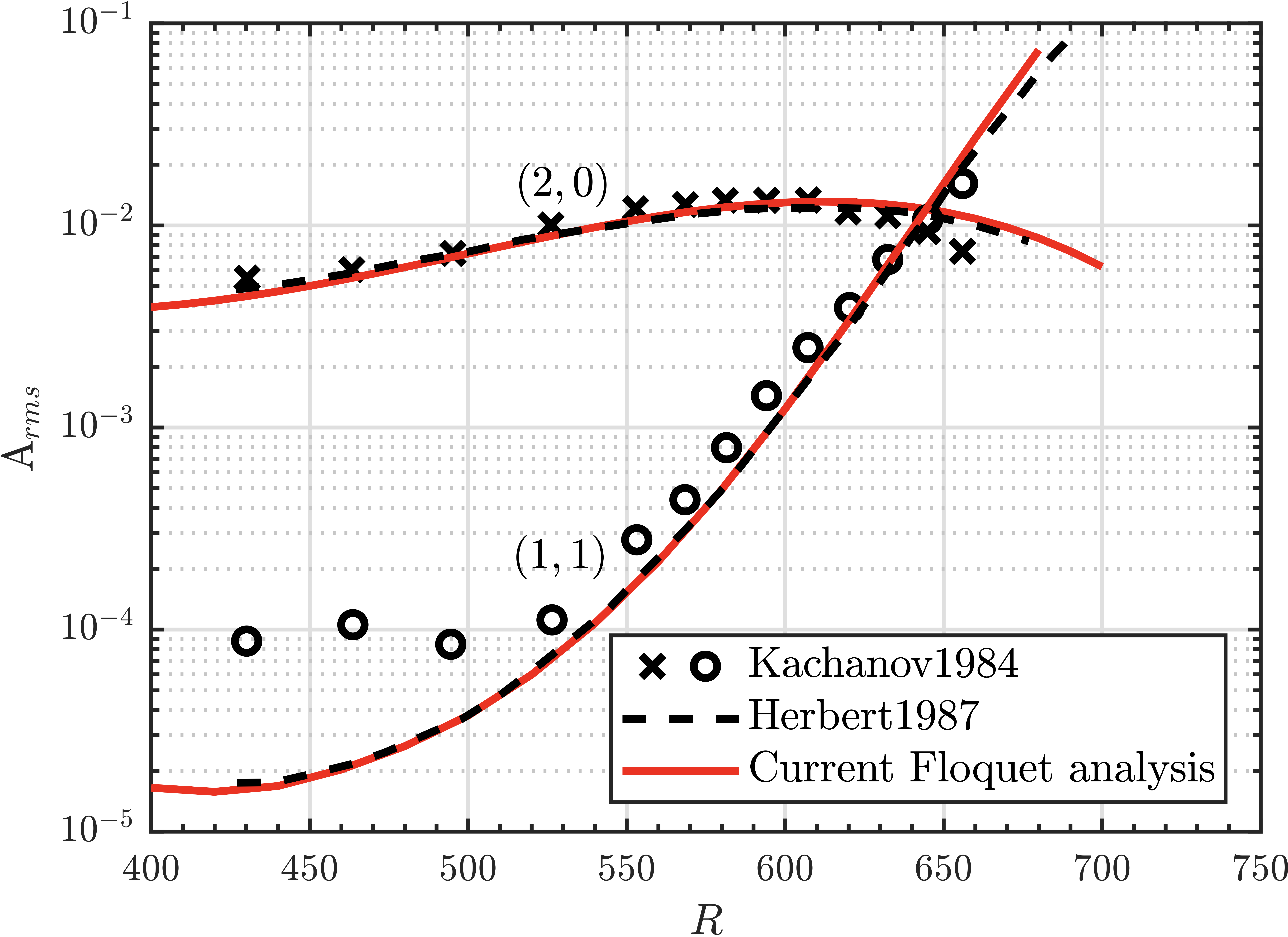} }
	\caption{Comparison among the current Floquet analysis, the Floquet analysis of \citet{Herbert1987} and the experimental data of \citet{Kachanov1984} for the (a) fundamental and (b) subharmonic modes at $R=600$, and (c) the amplitude growth of the two modes.}
	\label{fig:floquet_valid}
\end{figure}

The mode shape and the amplitude growth of the subharmonic mode are compared with the experimental data of \citet{Kachanov1984} and the Floquet analysis of \citet{Herbert1987}, as shown in figure~\ref{fig:floquet_valid}. The amplitude growth is obtained, using the integration 
	\begin{eqnarray}
		\frac{A_{(2,0)}(x)}{A_{(2,0)}(x_0)} &=& \int_{x_0}^{x} -\Im(\alpha_{(2,0)}) ds \\
		\frac{A_{(1,1)}(x)}{A_{(1,1)}(x_0)} &=& \int_{x_0}^{x} \Re(\gamma) ds = \int_{x_0}^{x} -\Im(\alpha_{(1,1)}) ds
	\end{eqnarray}
where the initial location $x_0$ corresponds to $R=400$ and the initial amplitudes are $A_{(2,0)}(x_0) =4\times 10^{-3}$ and $A_{(1,1)}(x_0) = 1.64 \times 10^{-5}$. The two mode shapes are identical to the literature data at $R=600$ in figure~\ref{fig:floquet_valid}a and \ref{fig:floquet_valid}b. The location $R=600$ is positioned in the subharmonic resonance range shown in figure~\ref{fig:floquet_valid}c. The amplitude growth from the current Floquet analysis matches well with the experimental data of \citet{Kachanov1984} and the Floquet analysis of \citet{Herbert1987}.  According to the discussion of \citet{Herbert1987}, the subharmonic mode in the experiment \citep{Kachanov1984} was not fully formed until $R \simeq 540$ because of the proximity to the vibrating ribbon for the disturbance generation. The measurement of \citet{Kachanov1984} showed that the disturbance mode $(1,1)$ had no definite phase value until $R \simeq 540$ and is deeply buried in background noise.  Note that the branch I of the neutral curve of the subharmonic mode in linear stability analysis is located near $R = 540$ \citep[see][figure 3]{Kachanov1984}.  It should be mentioned that the spatial growth data of the subharmonic mode \citep{Herbert1987} are used for the comparison here because the transformed data of the temporally-growing mode \citep{Herbert1984} overestimates the growth rate \citep{Herbert1987}.

The current Floquet analysis provides only the most resonating condition for the subharmonic mode for the given basic flow. Less resonating conditions, including the anti-resonant condition, can be obtained with the phase variation of the subharmonic oblique wave with the amplitude fixed. PSE and LES computations are explored for the investigation of the phase effect on the subharmonic resonance in the following sections.

\subsection{Phase effects on subharmonic resonance in PSE computations \label{sec:pse-only} }

Nonlinear parabolized stability equations (PSE) simulation is computationally efficient in investigating nonlinear interactions of instability modes in the transition region. The current PSE method, validated in the subharmonic resonance \citep{Park2013, Kim2019}, is explored here with the baseline inlet condition obtained from the current Floquet analysis. The inlet condition consists of the 2D fundamental TS and 3D subharmonic oblique modes. Further details of the baseline inlet condition, including the amplitude of the each mode, are provided in Section~\ref{sec:inlet}. 

The current Floquet analysis yields the phase difference between the fundamental and subharmonic modes $\Delta \phi_{in} = 130\degree$ at the inlet location $R=400$. Only the phase of the subharmonic mode varies so that the whole periodic range $0 \leq \Delta \phi_{in} \leq 180 \degree$ is simulated here. Other characteristics, including the amplitudes ($A_{(2,0)}$ and $A_{(1,1)}$) and the mode shapes ($\zeta$(y) and $\eta$(y)), remain the same at the inlet. 

\begin{figure}
	\centering
	\subfloat[]
	{\includegraphics[width=.5\textwidth]{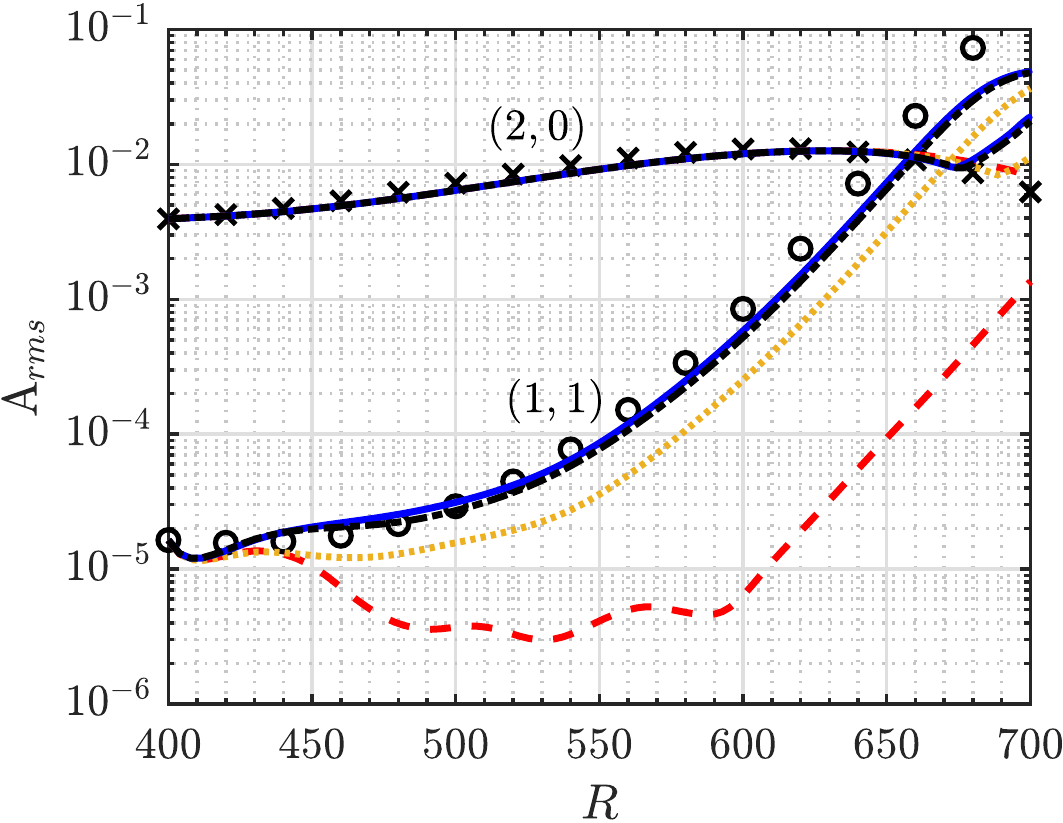}
	\label{fig:amplitude_growth_PSE} }
	\subfloat[]
	{\includegraphics[width=.5\textwidth]{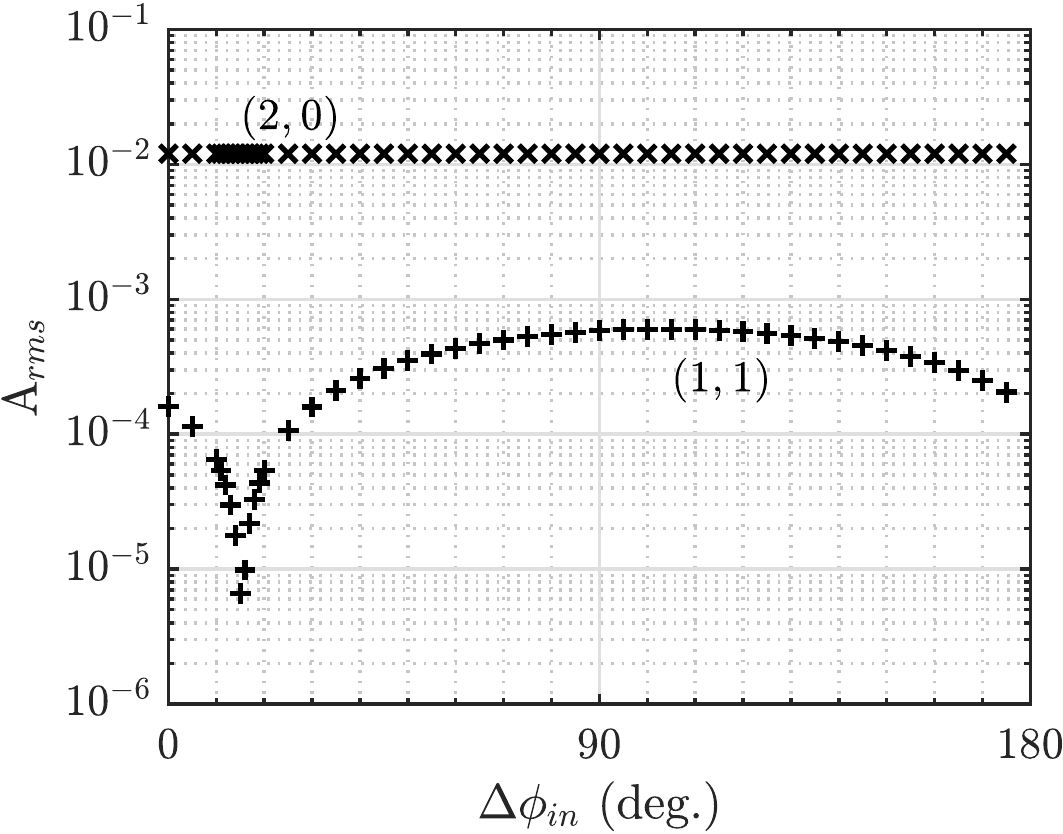}
	 \label{fig:amplitude_distribution_PSE}}
	\caption{
	Amplitude of the fundamental and subharmonic modes affected by the initial phase difference in the current simulation.
	(a) Amplitude growth: 
	$\times$,  Floquet analysis, (2,0); 
	\protect\markerone, Floquet analysis, (1,1); 
	\dashdot, PSE, $\Delta \phi_{in} = 130\degree$;  
	\blue{\full},  PSE, $\Delta \phi_{in} = 105\degree$; 
	\orange{\dotted}, PSE, $\Delta \phi_{in} = 40\degree$; 
	\red{\dashed}, PSE, $\Delta \phi_{in} = 15\degree$.  
	(b) Amplitude at $R=600$: 
	$\times$, PSE, (2,0); 
	$+$, PSE, (1,1).
	}
	\label{fig:amp_PSE}
\end{figure}

The initial phase difference significantly affects the amplitude evolution of the subharmonic mode as shown in figure~\ref{fig:amp_PSE}. Total 44 initial phases differences are simulated in the current PSE. Four selected phase differences, including the baseline $\Delta \phi_{in} = 130 \degree$, are shown in figure~\ref{fig:amp_PSE}a. The phase shift of 25 degrees from the baseline $\Delta \phi_{in} = 105 \degree$ still provides the almost identical evolution of the baseline amplitude for the subharmonic mode. The Floquet analysis data are similar to the PSE data of $\Delta \phi_{in} = 130 \degree$ and $\Delta \phi_{in} = 105 \degree$. The phase shift of 90 degrees from the baseline $\Delta \phi_{in} = 40 \degree$ yields a visual delay in the growth of the subharmonic mode. The case of  $\Delta \phi_{in} = 15 \degree$ even damps the subharmonic mode at the beginning until $R \simeq 600$, and the subharmonic mode starts to exponentially grow after $R=600$. The amplitude of the fundamental mode remains almost the same until $R \simeq 670$ regardless of the initial phase shift of the subharmonic mode. 

Figure~\ref{fig:amp_PSE}b shows the amplitude of the fundamental and subharmonic modes at $R=600$. The subharmonic mode resonates at most when the initial phase difference is $\Delta \phi_{in} = 105 \degree$. The subharmonic resonance is not sensitive to the phase if the phase difference is in the wide range of $60 \lesssim \Delta \phi_{in} \lesssim 150 \degree$ where the amplitude does not change more than a factor of 1.5.  In contrast, the subharmonic amplitude is significantly damped in the narrow range of $5 \lesssim \Delta \phi_{in} \lesssim 25 \degree$, more than a factor of 5 compared to the case of $\Delta \phi_{in} = 105 \degree$. At the phase valley $\Delta \phi_{in} = 15 \degree$, the subharmonic amplitude is even two-order-of-the-magnitude lower than that of $\Delta \phi_{in} = 105 \degree$. The least resonating condition is called anti-resonance. The phase difference between the most resonance and the anti-resonance in the current PSE is 90 degrees, as similarly observed in experiments with mild adverse pressure gradients \citep{Borodulin2002}. 

The Floquet analysis yields slightly large amplitude growth for the subharmonic mode, compared to the PSE analysis in the downstream region (see figure~\ref{fig:amplitude_growth_PSE}).  Since the disturbance equations are much simplified in the Floquet analysis, the detailed response of the subharmonic mode with respect to the given basic flow could be different to the PSE counterpart.  Additional PSE computations are conducted to investigate whether the assumptions used in the Floquet analysis contribute to the slight difference in the amplitude growth. Three distinct assumptions of the Floquet analysis are individually tested in the additional PSE computations: the parallel assumption of the basic flow, the exclusion of the nonlinear feedback from the subharmonic to fundamental mode, and the exclusion of the mean flow distortion (0,0) mode. Each individual assumption yields almost identical results to the PSE data shown in figure~\ref{fig:amplitude_growth_PSE}.  It can be conjectured that a different set of disturbance equations as a whole between the Floquet and PSE analysis may contribute to the subtle difference in the growth in figure~\ref{fig:amplitude_growth_PSE}.

The subharmonic resonance (secondary instability) is influenced by the fundamental mode (primary instability) via the nonlinear interaction. In contrast, almost no variation of the fundamental amplitude indicates that the feedback effect on the fundamental mode from the subharmonic mode is negligible until the subharmonic mode gains enough strength. In the current stability analysis using the Floquet and PSE computations, the early nonlinear region of the subharmonic resonance ends around $R=670$, and after that it can be expected that more modes rapidly evolve through higher nonlinear interactions (tertiary, quaternary, and so on). 

In the early nonlinear stage, it has been understood that the subharmonic mode resonates via the parametric resonance \citep{Herbert1987, El-Hady1988, Nayfeh1990}. Key parameters of the basic flow affecting the subharmonic resonance are the Reynolds number of the laminar flow $R$, the amplitude $A_{(2,0)}$ and the frequency $\omega$ of the fundamental mode, and the spanwise wavenumber $\beta $ of the subharmonic mode. In addition to these parameters, the phase difference between the two modes is also another important parameter, according to the current PSE study with various phase differences. Since the subharmonic resonance can be affected by the phase difference only, a complete transition to turbulent flow can be affected by the phase difference as well.  Because PSE becomes computationally expansive in the late nonlinear stage of the transition, Navier-Stokes equations are solved efficiently with the validated LES approach \citep{Jee2018, Kim2019, Kim2020, Lim2021} for the final transition to the turbulent flow, which is discussed in the next section.

\subsection{Phase effects on subharmonic resonance in LES computations \label{sec:les} }

High-fidelity large-eddy simulation (LES) is conducted here in order to study the effect of the phase difference between the two instabilities on both the subharmonic resonance and the complete turbulent transition. Total 9 LES cases are simulated with the inlet phase differences $\Delta \phi_{in} = 5, 7, 10, 11, 15, 20, 25, 105,$ and $130\degree$. The PSE analysis discussed in Section \ref{sec:pse-only} indicates that the subharmonic mode resonates for the given fundamental mode under the wide range of  $60 \lesssim \Delta \phi_{in} \lesssim 150 \degree$, so the two cases $\Delta \phi_{in} = 105$ and $130\degree$ are simulated here. The PSE analysis also suggests that the subharmonic resonance is significantly delayed when $5 \lesssim \Delta \phi_{in} \lesssim 25 \degree$, and the anti-resonant condition is highly sensitive to the initial phase. As a result, 7 phases $\Delta \phi_{in} = 5, 7, 10, 11, 15, 20,$ and $25\degree$ are selected in this narrow phase range. Note that the current LES approach has provided DNS-like fidelity for transitional boundary layer in authors' previous studies \citep{Kim2019, Kim2020}. LES statistics is obtained with LES data accumulated over 8 periods of the fundamental mode after two flow-through times in the LES domain. The time window for the statistics is larger here compared to the LES validation study of \citet{Kim2019} because anti-resonant conditions yield longer-time variations in the flow solution. 

\begin{figure}
	\centering
	\subfloat[Amplitude growth with $\Delta \phi _{in}=105\degree$]{
	\includegraphics[width=.72\textwidth]{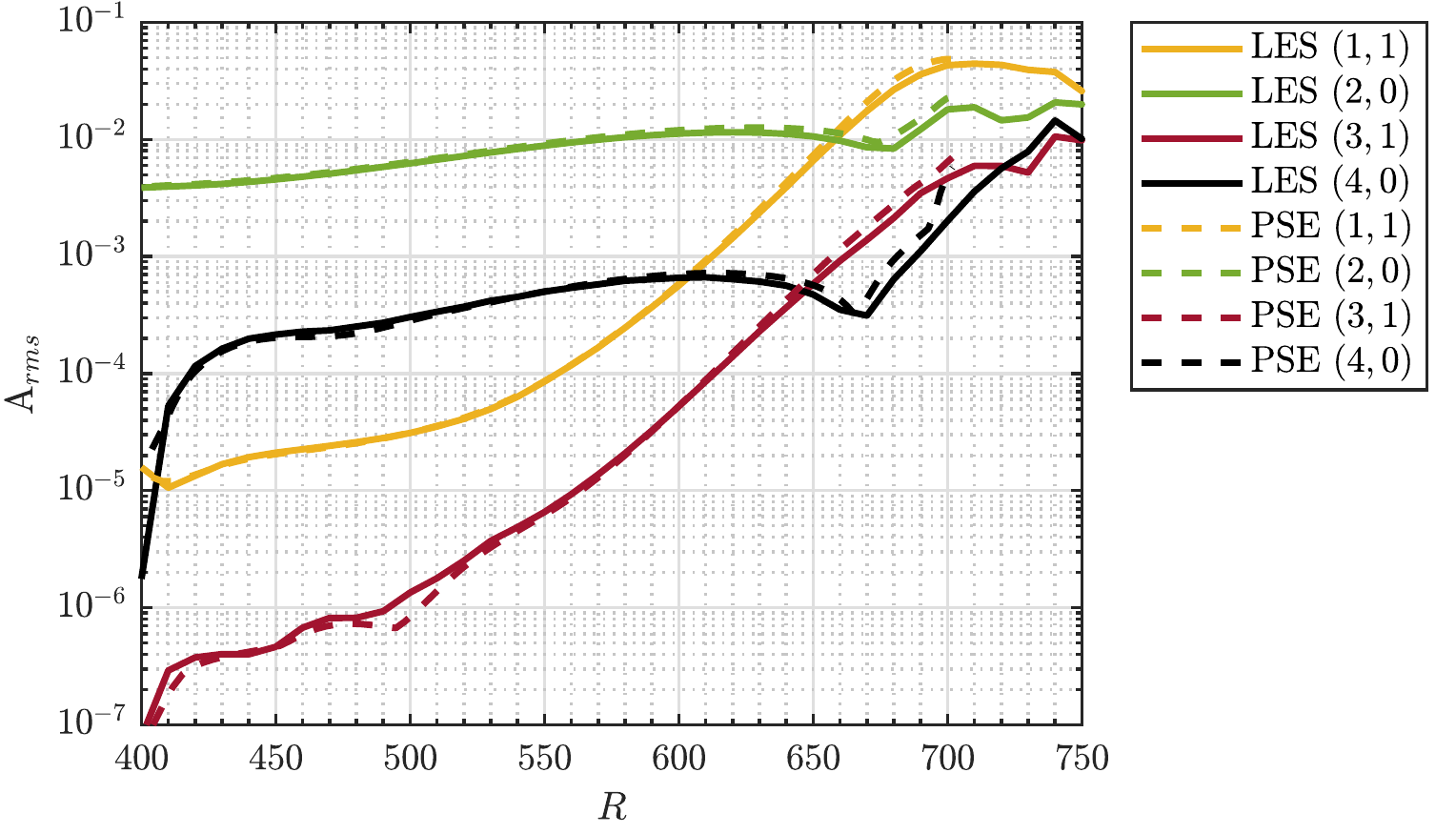} \label{fig:amplitude_growth_LES_105}
	}\\
	\subfloat[Amplitude growth with $\Delta \phi _{in}=15\degree$]{
	\includegraphics[width=.72\textwidth]{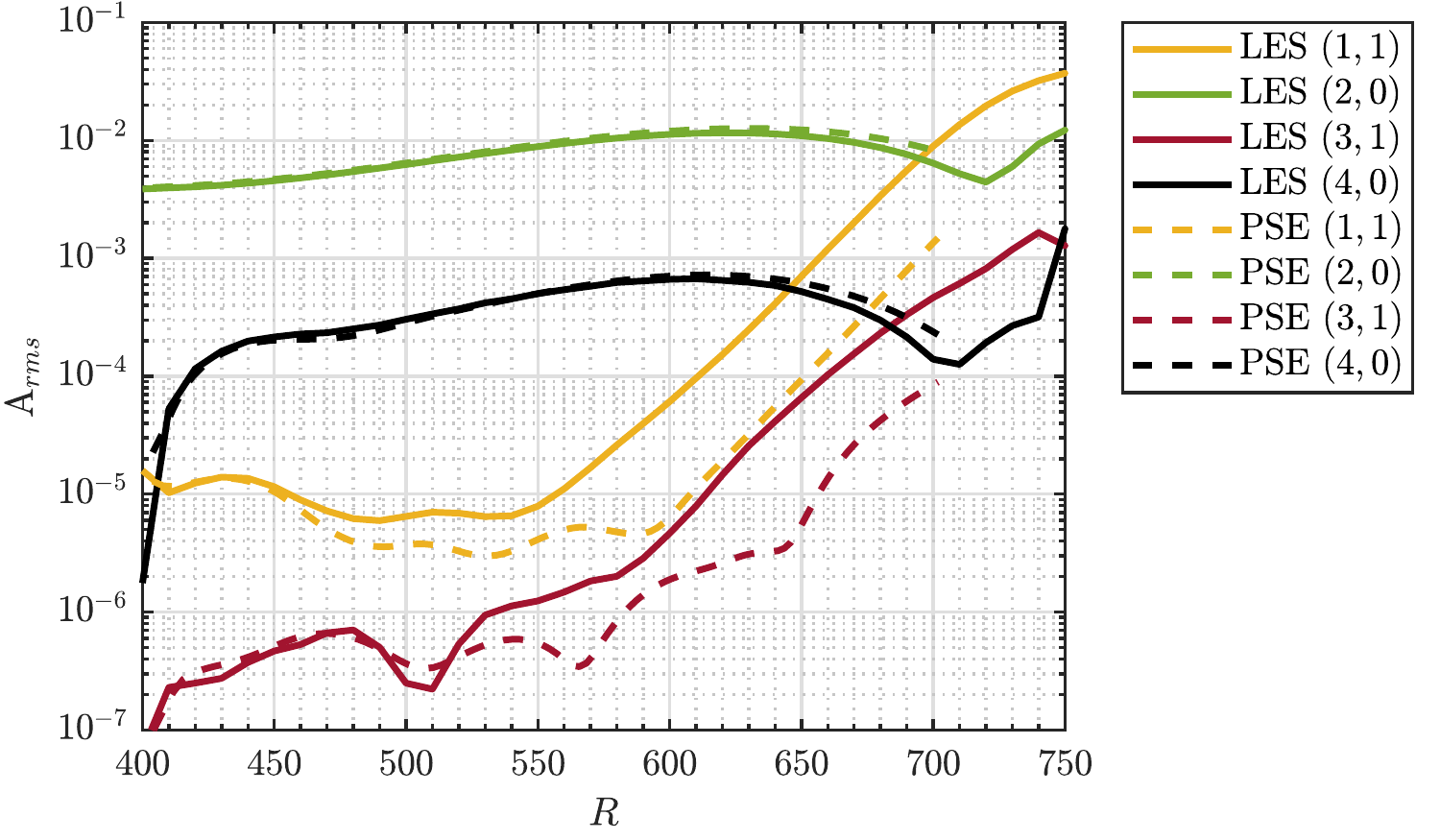} \label{fig:amplitude_growth_LES_15}
	}\\
	\subfloat[Amplitude growth with the anti-resonant phase]{
	\includegraphics[width=0.83\textwidth]{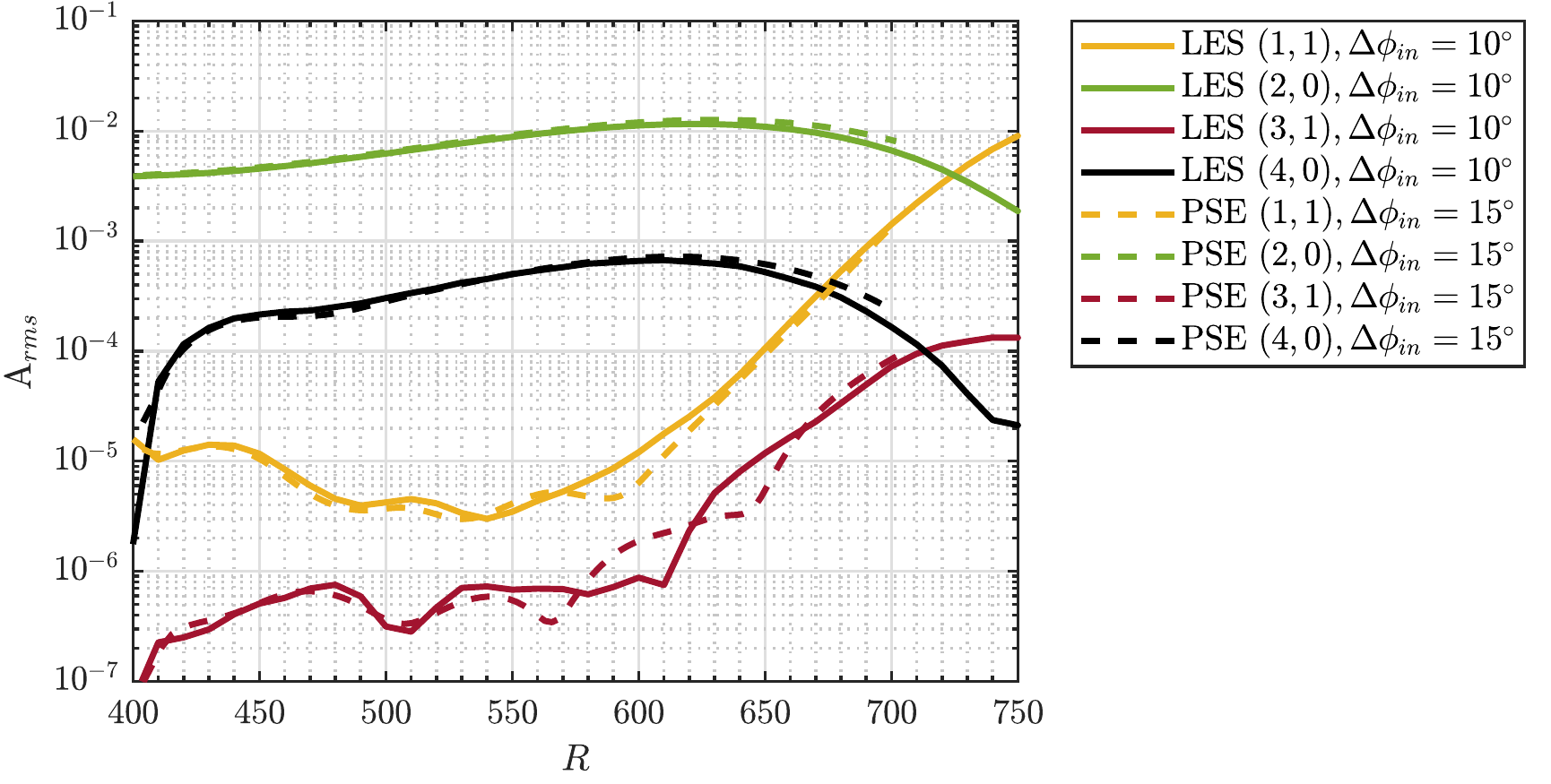} \label{fig:amplitude_growth_LES_10}
	}
	\caption{Amplitude growth of instability modes affected by the initial phase differences in the current PSE and LES computations}
	\label{fig:amplitude_growth_LES}
\end{figure}

The resonant and anti-resonant conditions are compared between the current LES and PSE computations, as shown in figure~\ref{fig:amplitude_growth_LES}. Two higher modes (3,1) and (4,0) are plotted along with the fundamental (2,0) and subharmonic (1,1) modes. The PSE is conducted until only $R=700$ owing to an expected surge in the computational cost for the downstream, late nonlinear transition stage. Instead, LES is conducted continuously in the further downstream up to $R=1050$. The amplitude growth in the LES computations is shown until $R=750$ in figure~\ref{fig:amplitude_growth_LES}.

Both LES and PSE provide almost the identical growth of selected instability modes in the resonant condition (see figure~\ref{fig:amplitude_growth_LES_105}). The resonant phase $\Delta \phi_{in} = 105 \degree$ yields the oblique modes (1,1) and (3,1) to grow exponentially from almost the inlet. Two planar modes (2,0) and (4,0) grow gradually until $R \simeq 650$ primarily owing to the linear growth of the fundamental mode (2,0). The harmonic mode (4,0) is mainly generated from the nonlinear effect of the fundamental mode, i.e., the nonlinear convective $\mathbb{N}$ terms in Eq.~\ref{eq:disturb}. After $R \simeq 670$, the transition undergoes a highly nonlinear stage, and all the modes grow exponentially in the current simulation. 

In figure~\ref{fig:amplitude_growth_LES_15}, the initial phase difference $\Delta \phi_{in} = 15 \degree$ is simulated in both LES and PSE. Both LES and PSE provide delayed subharmonic resonance. However, the subharmonic mode begins to resonate to the fundamental mode after $R \simeq 550$ in LES, whereas $R \simeq 600$ in PSE. In the current LES computations, $\Delta \phi_{in} = 10 \degree$ yields the least resonating condition for the subharmonic mode (anti-resonance), and the instability growth is very similar to the anti-resonant condition of PSE, as shown in figure~\ref{fig:amplitude_growth_LES_10}. The fundamental mode decays after $R \simeq 650$ owing to the insufficient growth of the subharmonic mode until $R=750$.  DNS computations (not shown here) in the anti-resonant condition also provide the almost identical results of the amplitude growth compared to LES, so the difference between LES and PSE for the anti-resonant phase may not come from the sub-grid-scale model which provides a negligible eddy viscosity until $R=700$.

\begin{figure}
	\centering
	\includegraphics[width=.65\textwidth]{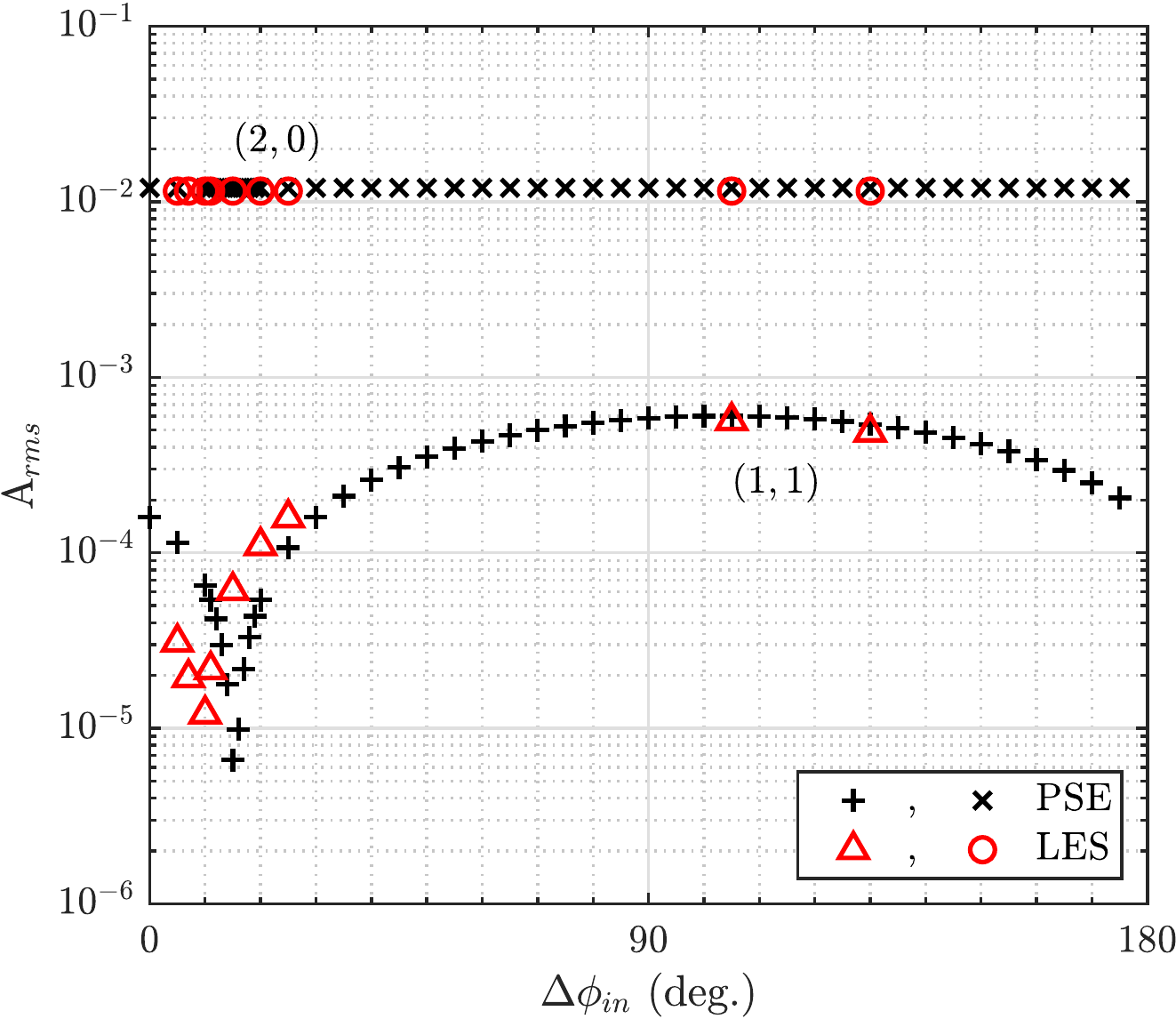}	
	\caption{Amplitude of the fundamental and subharmonic modes at $R=600$ affected by the initial phase difference in the current computations}
	\label{fig:amplitude_distribution_LES}
\end{figure}

The current LES and PSE computations indicate that the anti-resonance phenomena is highly sensitive to the flow condition as shown in figure~\ref{fig:amplitude_distribution_LES} where the amplitude of the fundamental and subharmonic modes at $R=600$ is plotted with the variation of the initial phase difference between the two modes. The anti-resonant phase is located in the narrow phase valley where the subharmonic amplitude varies about a factor of 10 only with the phase shift of 10 degrees. The anti-resonant phase is slightly different between LES and PSE; $\Delta \phi_{in} = 10 \degree$ in LES and  $\Delta \phi_{in} = 15 \degree$ in PSE. It is conjectured that the anti-resonance phenomena is sensitive to not only the initial phase difference but also the detailed flow solution of each computation.  Since the full Navier-Stokes equations (Eq.~\ref{eq:LES}) in LES are not identical to the parabolized equation of PSE, a subtle difference between LES and PSE computations can cause the slight difference for the anti-resonant phase. 

\begin{figure}
	\centering
	\includegraphics[width=\textwidth]{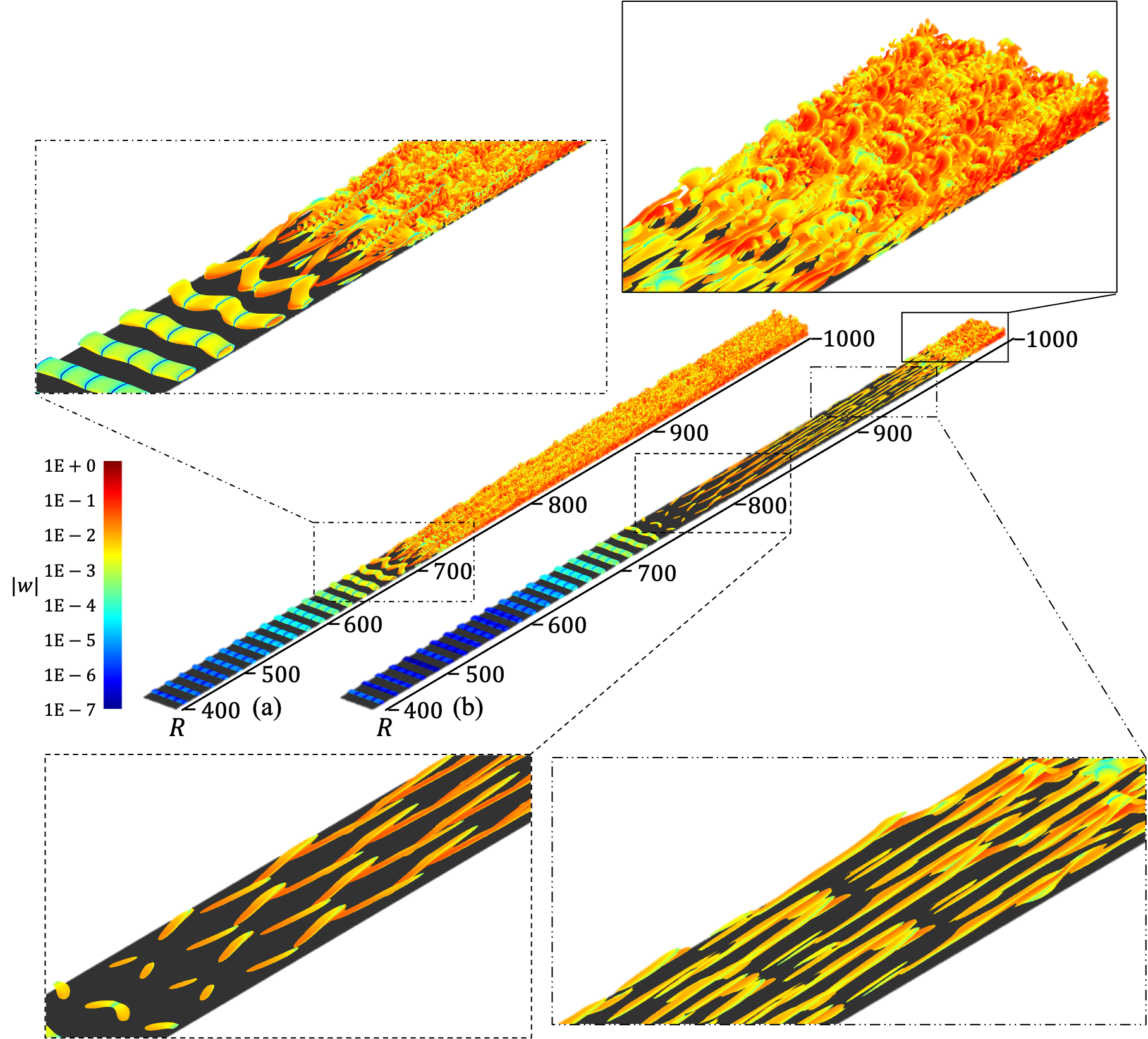}
	\caption{Vortical structures in the LES computations with the resonant phase $\Delta \phi _{in} = 105\degree$  (a) and the anti-resonant phase $\Delta \phi _{in}  = 10\degree$ (b). The iso-surface of the Q-criteria $Q=3 \tilde{U}_\infty^2 / \tilde{x}^2(R=400)$ is used for the visualization with the colour contour of the magnitude of the spanwise velocity $|w|$.}
	\label{fig:vortical_structure_LES}
\end{figure}

The phase effect on the subharmonic resonance eventually leads to a significant difference in the transition location to turbulent flow, as shown in figure~\ref{fig:vortical_structure_LES}. The resonant condition results in staggered $\Lambda$-shape vortical structures around $R=700$ which is the footprint of the subharmonic resonance observed in previous experiments \citep{Corke1989, Borodulin2011, Wuerz2012a} and high-fidelity computations \citep{Sayadi2013, Jee2018, Kim2019, Kim2020}. Fully turbulent flow starts roughly after $R=750$ in the resonant condition. In contrast, the anti-resonant condition leads to significantly delayed transition. Staggered $\Lambda$-shape vortical structures appear in the long range $ 750 \lesssim R \lesssim 950$ with prolonged structures near the end of the transition. The vortical structures are elongated in the streamwise direction probably because of the weak nonlinear interaction among low-amplitude instabilities expected from figure~\ref{fig:amplitude_growth_LES_10}.

\begin{figure}
	\centering
	\includegraphics[width=.77\textwidth]{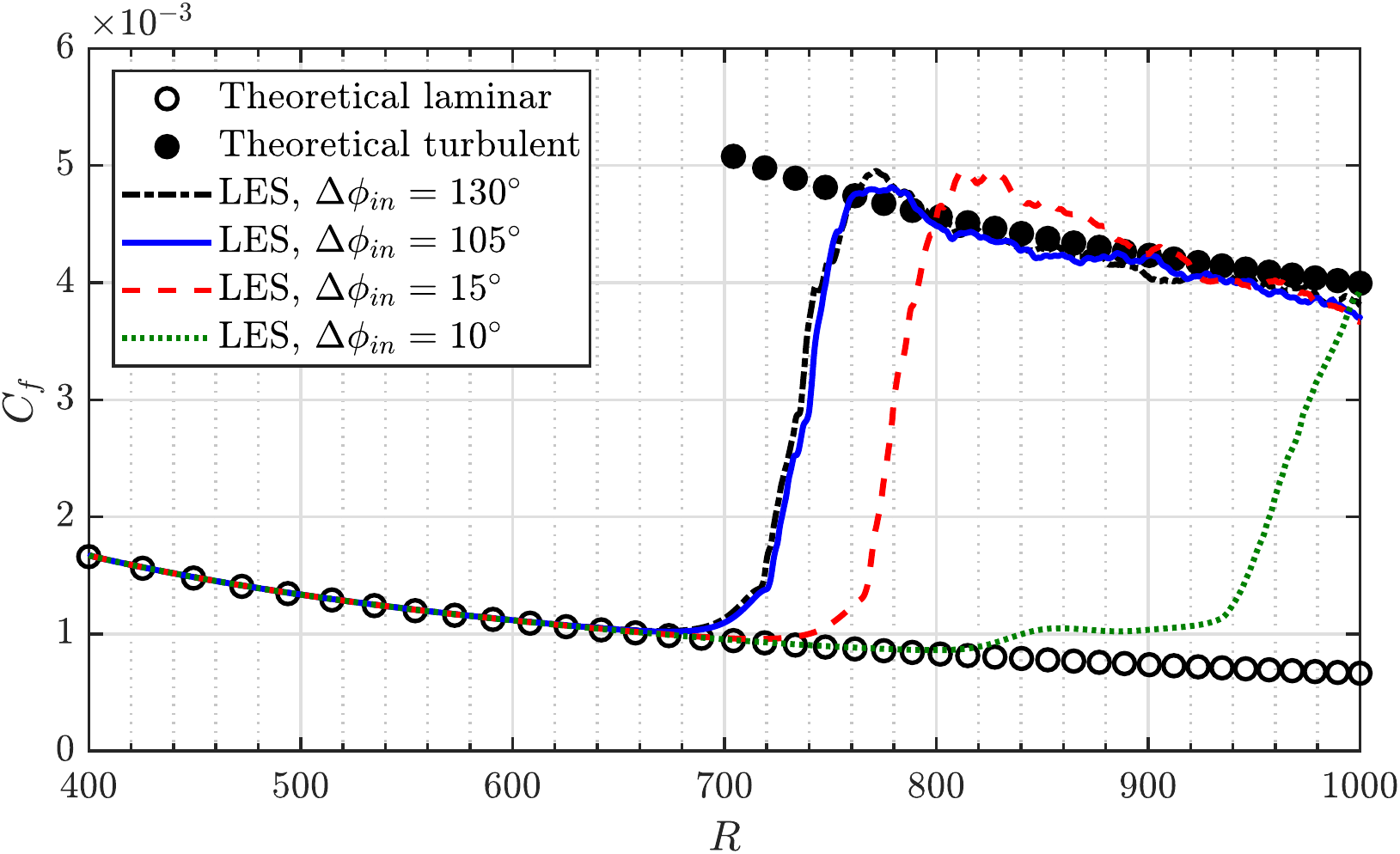}
	\caption{Skin friction $C_f$ in the current LES computations with various initial phase difference $\Delta \phi_{in}$, compared to theoretical data.}
	\label{fig:skin_friction_LES}
\end{figure}

The skin friction $C_f$ in figure~\ref{fig:skin_friction_LES} indicates the transition region affected by the initial phase difference. The resonant conditions ($\Delta \phi _{in} = 105$ and $130 \degree$) yield the deviation of $C_f$ from the laminar data around $R=690$ and the approach to the turbulent $C_f$ around $R=750$. In contrast, the anti-resonant condition with $\Delta \phi_{in}=10\degree$ leads to the $C_f$ deviation from the laminar around $R=820$ and the turbulent $C_f$ near $R=1000$. The prolonged vortical structures in figure~\ref{fig:vortical_structure_LES}b are associated with the long transition region. 

Current LES computations indicate that the initial phase difference by itself can control the transition location, which was not disclosed in the literature. The anti-resonant phase delays the turbulent transition location by $\Delta R \simeq 1000-750 = 250$ which corresponds to $\Delta Re_x \simeq 4.4 \times 10^5$, about 80\% increase in the transition Reynolds number from the resonant condition. The phase difference between the fundamental and subharmonic modes was modulated with an array of microphones in previous experiments with non-zero pressure gradient \citep{Borodulin2002, Wuerz2012}. Although complete transition to turbulent flow was not able to be achieved in the experiments of \citet{Borodulin2002, Wuerz2012}, the nonlinear interaction was significantly delayed with the anti-resonant phase. The current investigation and the experimental approach in controlling the phase suggest that the turbulent transition can be controlled with the phase modulation of a major instability mode (here the subharmonic mode).

\subsection{Discussion on resonance and anti-resonance \label{sec:evolution} }

\begin{figure}
	\centering
	\includegraphics[width=1\textwidth]{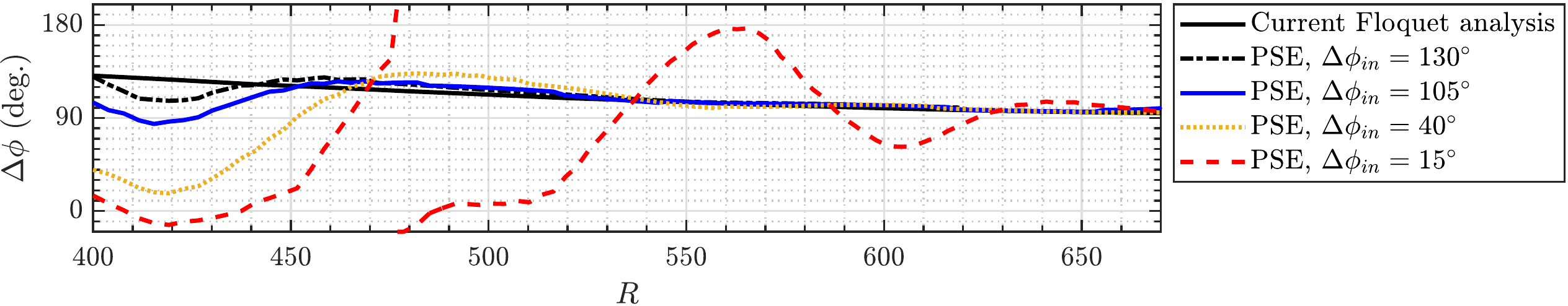}
	\caption{Evolution of the phase difference between the fundamental 2D and subharmonic 3D modes in the current PSE and Floquet analyses with the four selected initial phases.}
	\label{fig:evolution_phase_PSE}
\end{figure}

Resonance and anti-resonance phenomenon of the secondary instability (subharmonic mode) are further discussed here with the evolution of the phase difference between the fundamental and subharmonic modes. Total 44 initial phases differences are simulated in the current PSE (see figure~\ref{fig:amplitude_distribution_PSE}), and the evolution of four selected phases $\Delta \phi_{in}=15, 40, 105,$ and $130 \degree$ are shown in figure~\ref{fig:evolution_phase_PSE}. The Floquet analysis is also compared to the PSE data because the Floquet analysis provides the phase-locked condition. Initial phase differences near the Floquet $\Delta \phi_{in}$ follow the Floquet phase evolution in the current PSE with a slight deviation at the beginning. As the initial phase difference deviates further from the resonant phase, the phase evolution requires more distance to approach the Floquet phase difference.  It takes about $\Delta R=200$ for the anti-resonant condition of the initial phase $\Delta \phi_{in} = 15\degree$ to catch up with the subharmonic resonance, which is consistent with the exponential growth of the subharmonic mode after $R\simeq 600$ in figure~\ref{fig:amplitude_growth_PSE}.

\begin{figure}
	\centering
	\includegraphics[width=1\textwidth]{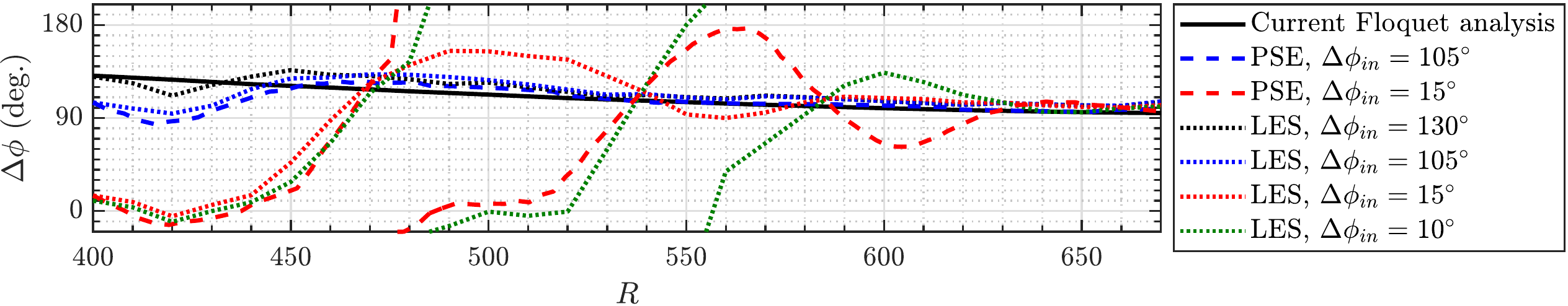}
	\caption{Evolution of the phase difference between the fundamental 2D and subharmonic 3D modes in the current PSE, LES and Floquet analyses with the four selected initial phases.}
	\label{fig:evolution_phase_LES}
\end{figure}

The phase evolution is also confirmed in the LES computations as shown in figure~\ref{fig:evolution_phase_LES}. Two resonant initial phase differences ($\Delta \phi_{in}=105$ and $130 \degree$) and two anti-resonant phase differences ($\Delta \phi_{in}=10$ and $15 \degree$) are selected for the LES data sets. Two selected PSE cases are also plotted for the comparison. In the resonant conditions, the phase difference quickly converges to the Floquet phase difference. For the anti-resonant condition of $\Delta \phi_{in}=10 \degree$ in LES, the phase difference approaches the Floquet data after $R=600$, similarly observed in the PSE anti-resonance of $\Delta \phi_{in}=15 \degree$. Note that the anti-resonance phenomena is sensitive to detailed numerical solution, so the difference of 5 degree seems acceptable, as discussed in Section \ref{sec:les}. 

In both PSE and LES, regardless of the initial phase differences, $\Delta \phi$ converges to the resonant phase difference about 90 degrees in the downstream. Similar evolution of the phase difference between two major instability modes (normally primary and secondary instability modes) has been observed in experiments \citep{Borodulin2002, Wuerz2012}. \citet{Borodulin2002} noticed the similar convergence to 90 degree for the phase difference in adverse-pressure-gradient boundary layer flow on a flat plate \citep[see][figure 26]{Borodulin2002}. The narrow phase range for the anti-resonant condition was also observed in the experiment \citep[see][figure 25]{Borodulin2002}. \citet{Wuerz2012} also measured the evolution of the phase difference starting from the anti-resonance to the resonance in a boundary layer on a laminar airfoil. 

The mechanism of the phase evolution is related to the phase synchronization of the subharmonic mode for the parametric resonance. A simple dynamic system expressed in Mathieu's equation, which describes a sinusoidal parametric excitation \citep{Kovacic2018}, requires the phase synchronization for the parametric resonance. An initial phase shift of the subharmonic mode eventually approaches to the phase synchronization \citep{Kim2020a}, and the transient interval is associated with the initial phase shift. \citet{Kim2020a} obtained two opposite local solutions, one for exponential growth (resonance) and another for exponential damping (anti-resonance) in Mathieu's equation. An arbitrary initial condition can be decomposed into these two solutions. Obviously, the resonant component grows exponentially and dominates the subharmonic mode, which results in the phase evolution to the phase synchronization. A similar discussion can be found in the experimental observation on the phase evolution \citep{Borodulin2002, Wuerz2012}.

\begin{figure}
	\centering
	\includegraphics[width=.85\textwidth]{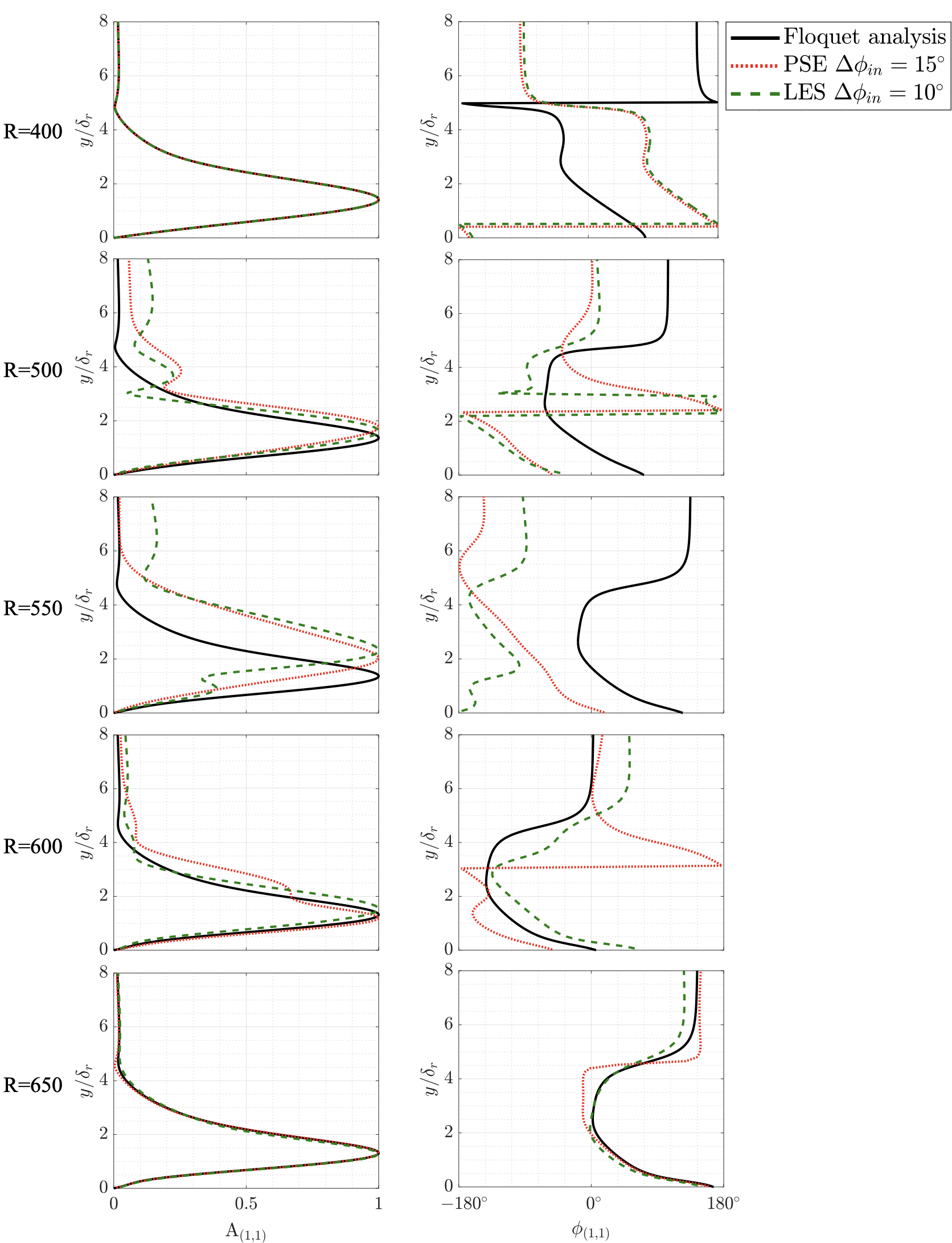}		
	\caption{Comparison of the modal shapes between the current Floquet analysis, PSE, and LES computations for anti-resonant conditions. The amplitude is scaled by its own maximum in each case.}
	\label{fig:mode_shape_anti-resonance}
\end{figure}

\begin{figure}
	\centering
	\includegraphics[width=.86\textwidth]{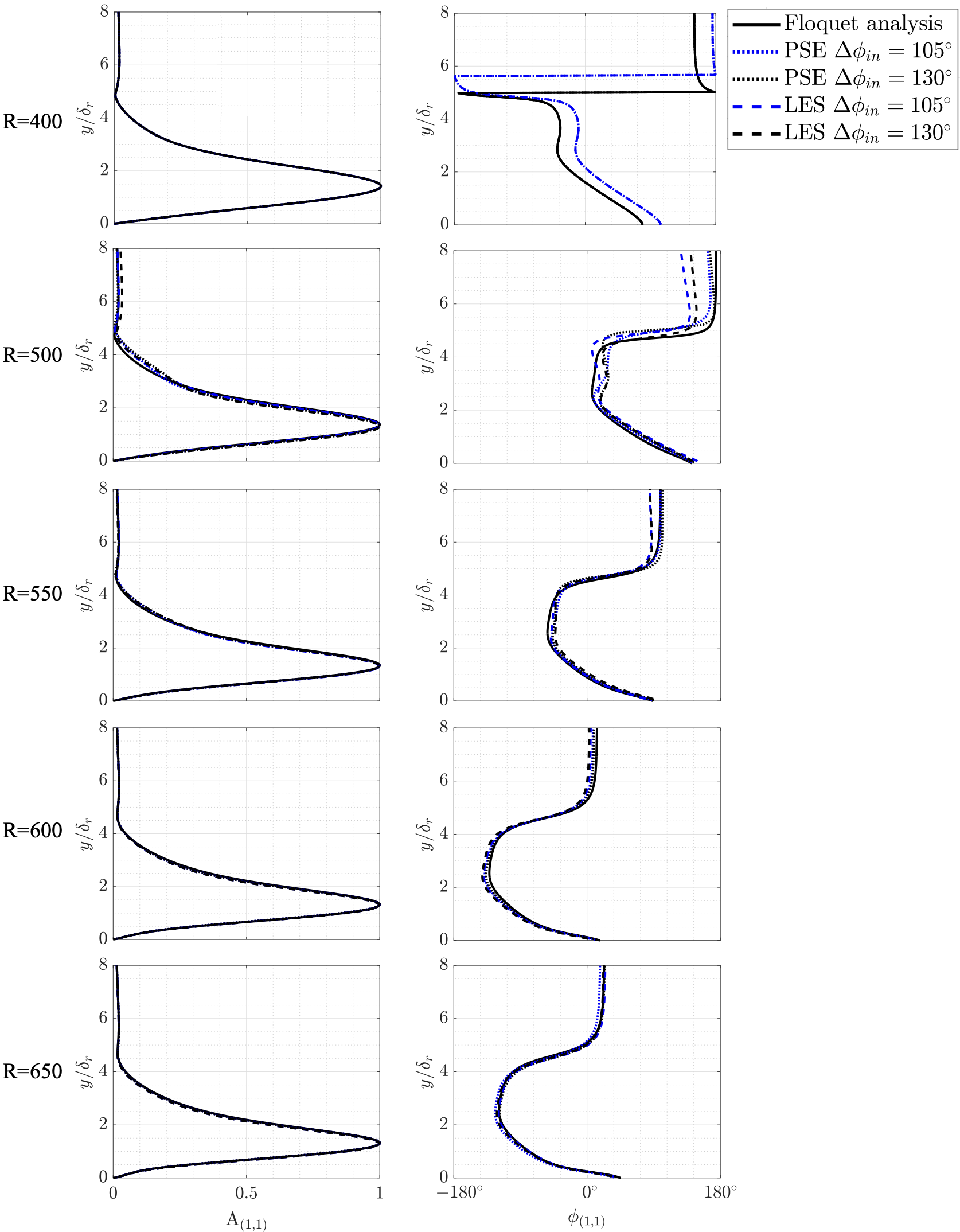}		
	\caption{Comparison of the modal shapes between the current Floquet analysis, PSE, and LES computations for resonant conditions. The amplitude is scaled by its own maximum in each case.}
	\label{fig:mode_shape_resonance}
\end{figure}	

Initial phase differences in the anti-resonant condition change the mode shape of the subharmonic mode during the transient region towards the phase synchronization, as shown in figure~\ref{fig:mode_shape_anti-resonance}. In the initial damping region $400 < R < 600$ of the anti-resonant condition, the subharmonic mode undergoes a severe distortion. As the phase difference approaches to the resonant phase predicted by the Floquet analysis around $R=600$ (i.e., the phase synchronization), the subharmonic mode eventually recovers to the resonant shape. In contrast, the resonant condition maintains the initial resonant modal shape as shown in figure~\ref{fig:mode_shape_anti-resonance}. Note that the resonant condition is insensitive to the phase shift if the phase difference is in the wide range of $60 \lesssim \Delta \phi_{in} \lesssim 150 \degree$, which has been discussed with figure~\ref{fig:amplitude_distribution_PSE}.

\clearpage

\section{Conclusions}\label{sec:conclusions}

A thorough investigation was conducted here for the phase effect on the growth of the secondary instability in the early nonlinear transition region and its final impact on the turbulent transition location. Numerical methods with various levels of fidelity were systematically explored in this work. The fundamental 2D TS wave and subharmonic oblique wave were chosen for the primary and corresponding secondary instability, respectively, in the current study.

The Floquet analysis provides a resonating subharmonic mode for a given basic flow composed of steady laminar flow and the fundamental TS wave with the assumption of locally parallel basic flow and one-way nonlinear effect from the basic flow to subharmonic mode. The current Floquet analysis was validated against the pioneered analysis of \citet{Herbert1987} and the experimental data of \citet{Kachanov1984} for the same flow condition.  In such an early nonlinear stage in the laminar region, the local Floquet analysis is efficient for obtaining the subharmonic mode resonating to the given basic flow. 

Because the current Floquet analysis is limited to a resonating condition, PSE was used for the new parametric study on the effect of the phase on the subharmonic resonance. Total 44 phase differences between the fundamental and subharmonic modes were studied by shifting the subharmonic phase with respect to the given fundamental TS wave while maintaining the initial amplitude of both the waves at the PSE inlet. The current PSE indicates that the subharmonic resonance is insensitive to the phase difference as long as the phase shift is less than about $45 \degree$ from the most resonating phase condition. In contrast to the resonance phenomena, the anti-resonant condition damps the subharmonic growth at the beginning and delays the resonance significantly. The anti-resonant condition is sensitive to the phase difference here --- only $5 \degree$ difference can cause the subharmonic amplitude to vary with a factor of 10 downstream near $R=600$ in the current simulation. 

To carry through complete simulations to fully turbulent flows, LES was used for 9 selected phase differences (2 in the resonant condition and 7 near the anti-resonance). The current LES provides the subharmonic resonance observed in the Floquet and PSE analyses in the early nonlinear region, indicating that LES is essentially DNS before highly nonlinear interactions occur. PSE and LES were also compared to each other in less-resonating conditions, and the comparison is acceptable even in the anti-resonant condition. A slight difference, about $5 \degree$ in the anti-resonant phase between PSE and LES, is presumably related to the sensitive nature of the anti-resonance phenomena to detailed flow solutions. The turbulent transition is significantly delayed when the subharmonic resonance is mostly suppressed in the anti-resonant condition. The variation of the transitional Reynolds number can be drastic as $ \Delta Re_{x, tr} \simeq 4.4\times 10^5$, when the phase shifts from the resonant to anti-resonant condition. Although a realistic transition control technique with the phase modulation requires further investigation, a few experimental approaches with an array of microphones in \citet{Borodulin2002, Wuerz2012} indicate the control potential. 

The resonance and anti-resonance phenomenon were further discussed with the evolution of the phase difference. In a resonating condition, the phase difference quickly follows the phase evolution of the Floquet analysis even though the initial phase is moderately deviated from the Floquet phase. In the anti-resonant condition, it takes a significant distance for the subharmonic mode to catch up on the phase evolution of the Floquet analysis. The mechanism of the phase evolution is associated with the phase synchronization of the subharmonic mode for the parametric resonance. A similar phase evolution towards the resonant phase has been noticed even in a simple nonlinear dynamic system governed by Mathieu's equation \citep{Kim2020a} and experimental measurements under non-zero pressure gradients \citep{Borodulin2002, Wuerz2012}. In the anti-resonant condition, the subharmonic mode shape deviates from the resonating shape at the beginning as the subharmonic amplitude damps, and then it returns back to the resonating shape as the phase synchronization occurs.

\section*{Acknowledgments}	
This work was supported by the National Research Foundation of Korea (NRF) grant funded by the Korea government (MSIT) (Project No. NRF-2021R1A2C1006193) and Space Core Technology Development Program (No. NRF-2017M1A3A3A02016810). Computational resources were supported by the KISTI National Supercomputing Center (Project Numbers: KSC-2020-CRE-0057 and KSC-2018-CRE-0113) with technical support. 

\section*{Declaration of Interests}
The authors report no conflict of interest.

\appendix
\section{Primary and Secondary Instability (Floquet) Analyses \label{appendix}}
\numberwithin{equation}{section}
\setcounter{equation}{0}

	Equations for the primary instability are obtained for the fundamental TS wave, substituting Eqs. \ref{eq:2D_basic} and \ref{eq:2D_dist} in Eq. \ref{eq:disturb}, and eliminating nonparallel terms, as provided in Appendix of \citet{El-Hady1988},
	\begin{subeqnarray}
	 \dot{\zeta}_1 &=& \zeta_2 \\
	 \dot{\zeta}_2 &=& \left\{ \alpha^2 + i R_o (\alpha U_L-\omega) \right \} \zeta_1 + R_o \dot{U}_L \zeta_3 +i \alpha R_o \zeta_4 \\
	 \dot{\zeta}_3 &=& -i\alpha \zeta_1 \\
	 \dot{\zeta}_4 &=& -\frac{i \alpha \zeta_2}{R_o} - \left\{ \frac{\alpha^2}{R_o} + i (\alpha U_L-\omega) \right\} \zeta_3
	 \label{eq:Pri}
	\end{subeqnarray}
	where $\dot{\zeta}=d\zeta/dy$. The set of the first-order differential equations with the boundary conditions given in Eq. \ref{eq:Pri_BC} yields an eigenvalue problem for the eigenvalue $\alpha$ and the eigenfunctions $\zeta$'s.
	\begin{subeqnarray}
	\zeta_1 =\zeta_3 &=& 0 \quad \text{at} \quad y=0 \\ 
	\zeta_1,\zeta_3 &\to& 0 \quad \text{as} \quad y \to \infty
	\label{eq:Pri_BC}
	\end{subeqnarray}

	Substituting Eqs. \ref{eq:3D_basic} and \ref{eq:3D_dist} for the basic flow and the disturbance in Eq. \ref{eq:disturb}, respectively, with $\{u, v, w, p \}$ being replaced with $\{u, v, w, p \}_{1/2}$, and equating the coefficients of $\exp(\pm i\theta/2)$ on both sides, the following set of first-order ordinary differential equations for the secondary instability is obtained.
	
	\begin{subeqnarray} 
	 \dot{\eta}_1 &=& \eta_2  \\
	 \dot{\eta}_2 &=& \Gamma\eta_1 + R_o \dot{U}_L \eta_3 + R_o(\gamma + i\alpha_{1/2}) \eta_4  \nonumber \\
	 & & + AR_o(\gamma + i\alpha_{1/2})\zeta_1 \eta_7 + AR_o\zeta_3 \eta_8 + AR_o\zeta_2 \eta_9  \\
	 \dot{\eta}_3 &=& -(\gamma + i\alpha_{1/2})\eta_1 -\beta \eta_5  \\
	 \dot{\eta}_4 &=& -\frac{1}{R_o}[(\gamma + i\alpha_{1/2})\eta_2 + \Gamma\eta_3 + \beta \eta_6] \nonumber \\
	 & & +A(\gamma - i3\alpha_{1/2})\zeta_3 \eta_7 -A[\gamma - i(\alpha+\alpha_{1/2})]\zeta_1\eta_9 +A\beta \zeta_3 \eta_{11}  \\
	 \dot{\eta}_5 &=& \eta_6  \\
	 \dot{\eta}_6 &=& -\beta R_o \eta_4 + \Gamma \eta_5 +AR_o(\gamma - i\alpha_{1/2})\zeta_1 \eta_{11} + AR_o \zeta_3 \eta_{12} \\
	 \dot{\eta}_7 &=& \eta_8  \\
	 \dot{\eta}_8 &=& \Gamma \eta_7 + R_o \dot{U}_L \eta_9 + R_o(\gamma - i\alpha_{1/2}) \eta_{10} \nonumber \\
	 & & +AR_o(\gamma - i\alpha_{1/2}) \zeta^*_1 \eta_1+ AR_o \zeta^*_3 \eta_2 + AR_o \zeta^*_2 \eta_3  \\
	 \dot{\eta}_9 &=& -(\gamma - i\alpha_{1/2})\eta_7 -\beta \eta_{11}  \\
	 \dot{\eta}_{10} &=& -\frac{1}{R_o}[(\gamma - i\alpha_{1/2})\eta_8 + \Gamma^* \eta_9 + \beta \eta_{11}]   \nonumber \\
	 & & +A(\gamma + i3\alpha_{1/2})\zeta^*_3 \eta_1 -A[\gamma + i(\alpha^*+\alpha_{1/2})]\zeta^*_1\eta_3 +A\beta \zeta^*_3 \eta_5  \\
	 \dot{\eta}_{11} &=& \eta_{12}  \\
	 \dot{\eta}_{12} &=& -\beta R_0 \eta_{10} + \Gamma^* \eta_{11} +AR_o(\gamma + i\alpha_{1/2})\zeta^*_1 \eta_5 + AR_o \zeta^*_3 \eta_6 
	 \label{eq:Sec}
	\end{subeqnarray}
	where $\Gamma=\alpha_{1/2}^2+\beta^2-\gamma^2 - 2i\alpha_{1/2}\gamma + R_o[\gamma U_L+i(\alpha_{1/2} U_L -\omega_{1/2}))] $ and $\beta$ is the spanwise wavenumber for the subharmonic wave. Note that the presence of the fundamental TS wave in the basic flow provides the nonlinear interaction between the fundamental and subharmonic modes through all the terms including $U$ and $V$ in Eq. \ref{eq:disturb} (even the square bracket terms).  The nonlinear interaction is one way in the current analysis because only the subharmonic oblique wave is influenced by the fundamental TS wave not vice versa.  In the derivation from Eq. \ref{eq:disturb} to Eq. \ref{eq:Sec}, higher subharmonic modes such as $\exp(i3\theta/2)$ are ignored. 
	
	Note that the current formulation in Eq. \ref{eq:3D_dist} is an extended version of \citet[Eq. 33]{Nayfeh1990} and \citet[Eq. 14]{El-Hady1988} in order to cooperate the complex exponent $\gamma$ later, whereas the real $\gamma$ is assumed in \citet{Nayfeh1990, El-Hady1988}. Although the current Floquet analysis is able to handle the complex $\gamma$, the subharmonic oblique wave perfectly synchronized with the fundamental TS wave, i.e., the real $\gamma$, is of interest here.  The physical meaning of the function $\eta$ is given in table \ref{tab:psi} when $\gamma$ is real. 

	\begin{table}
	\centering
	\begin{tabular}{c c | c c} 
	Function  	& Physical Meaning 					& Function 		& When $\gamma=\Re(\gamma)$ \\ 
	$\eta_1$ 	& modal shape of $u_{1/2} $ 			& $\eta_7$ 		& $\eta_1^*$ \\ 
	$\eta_2$ 	& modal shape of $\dot{u}_{1/2} $ 	& $\eta_8$ 		& $\eta_2^*$ \\
	$\eta_3$ 	& modal shape of $v_{1/2} $  			& $\eta_9$ 		& $\eta_3^*$ \\
	$\eta_4$ 	& modal shape of $p_{1/2} $  			& $\eta_{10}$ 	& $\eta_4^*$ \\
	$\eta_5$ 	& modal shape of $w_{1/2} $  		& $\eta_{11}$ 	& $\eta_5^*$ \\
	$\eta_6$ 	& modal shape of $\dot{w}_{1/2} $ 	& $\eta_{12}$ 	& $\eta_6^*$ \\
	\end{tabular}
	\caption{Physical meaning of function $\eta$ for indices 1-6 and the relevant additional function when $\gamma$ is real}
	\label{tab:psi}
	\end{table}	

	The set of the twelve differential equations (Eq. \ref{eq:Sec}) with boundary conditions given in Eq. \ref{eq:Sec_BC} yields an eigenvalue problem for the eigenvalue \(\gamma\) and the eigenfunction \(\eta\).
	\begin{subeqnarray} \label{eq:Sec_BC}
	\eta_1 =\eta_3 =\eta_5 =\eta_7 =\eta_9 =\eta_{11} = 0 &\text{\quad at \quad}& y=0 \\
	\eta_1 ,\eta_3 ,\eta_5 ,\eta_7 ,\eta_9 ,\eta_{11} \rightarrow 0 &\text{\quad as \quad}& y \rightarrow \infty
	\end{subeqnarray}
	
	The Blasius solution is used for the laminar flow $U_L(y)$. Although the Blasius solution is not strictly parallel (i.e., $V \neq 0$ and $\partial U_L/\partial x \neq 0$), the assumption of the locally parallel flow is used for the analysis of both the primary instability (Eq. \ref{eq:Pri}) and the secondary subharmonic instability (Eq. \ref{eq:Sec}), following the previous approach in \citet{Herbert1984,Herbert1987,Nayfeh1990,El-Hady1988}. 
	
	The total number of the grid points is 512 in the wall-normal distance of $200 \tilde{\delta}_r$ with about 300 points inside the boundary layer thickness $5 \tilde{\delta}_r$, and the first wall-normal grid size is $7\times 10^{-3} \tilde{\delta}_r$. The current wall-normal mesh points are obtained using the mapping from a semi-infinite to a finite domain suggested in \citet[Appendix A.4]{Schmid2001} as described in Eq. \ref{eq:Mesh}. 
	\begin{eqnarray}
	& y_j = c_1 \frac{1+h_j}{c_2-h_j}, \quad h_j = \frac{2j - N}{N}, \quad j=0, 1, \cdots, N, \quad N=512  \label{eq:Mesh} \\
	& y_0 = 0, \quad y_{N/2}=\frac{c_1}{c_2} = 3.5, \quad y_N = \frac{c_1}{c_2-1}=200, \quad c_1=\frac{700}{193}, \quad c_2=\frac{196.5}{193} \nonumber
	\end{eqnarray}
		
	Using the Matlab built-in function {\it polyeig}, the eigenvalue problem with Eqs. \ref{eq:Pri} and \ref{eq:Pri_BC} is solved for the 2D TS wave. The same function {\it polyeig} is used for the Floquet analysis (Eq. \ref{eq:Sec}) with the boundary condition (Eq. \ref{eq:Sec_BC}) to obtain the subharmonic characteristic exponent $\gamma$ along with the eigenfunctions $\eta$'s.  The function {\it polyeig} solves the polynomial eigenvalue problem $(\mathsfbi{A}_0 + \mathsfbi{A}_1 \lambda + \cdots + \mathsfbi{A}_n \lambda^n) \boldsymbol{b} = 0$ where $\lambda$ is the eigenvalue, $\boldsymbol{b}$ is the eigenvector and $\mathsfbi{A}$ is the given square coefficient matrix \citep[see][Ch. 9.8]{Higham2016}. For the current quadratic eigenvalue problems, $n=2$.


\begin{thebibliography}{35}
\expandafter\ifx\csname natexlab\endcsname\relax\def\natexlab#1{#1}\fi
\def\au#1{#1} \def\ed#1{#1} \def\yr#1{#1}\def\at#1{#1}\def\jt#1{\textit{#1}}
  \def\bt#1{#1}\def\bvol#1{\textbf{#1}} \def\vol#1{#1} \def\pg#1{#1}
  \def\publ#1{#1}\def\arxiv#1{#1}\def\org#1{#1}\def\st#1{\textit{#1}}

\bibitem[Bertolotti {\em et~al.\/}(1992)Bertolotti, Herbert \&
  Spalart]{Bertolotti1992}
{\sc \au{Bertolotti, F.~P.}, \au{Herbert, T.} \& \au{Spalart, P.~R.}} \yr{1992}
   \at{Linear and nonlinear stability of the {B}lasius boundary layer}.
  \jt{Journal of Fluid Mechanics}  \bvol{242},  \pg{441--474}.

\bibitem[Borodulin {\em et~al.\/}(2002{\natexlab{{\em a\/}}})Borodulin,
  Kachanov \& Koptsev]{Borodulin2002}
{\sc \au{Borodulin, V.~I.}, \au{Kachanov, Y.~S.} \& \au{Koptsev, D.~B.}}
  \yr{2002}  \at{Experimental study of resonant
  interactions of instability waves in a self-similar boundary layer with an
  adverse pressure gradient: {I. Tuned} resonances}.  \jt{Journal of
  Turbulence}  \bvol{3},  \pg{N62},  \arxiv{arXiv:
  https://doi.org/10.1088/1468-5248/3/1/062}.

\bibitem[Borodulin {\em et~al.\/}(2011)Borodulin, Kachanov \&
  Roschektayev]{Borodulin2011}
{\sc \au{Borodulin, V.~I.}, \au{Kachanov, Y.~S.} \& \au{Roschektayev, A.~P.}}
  \yr{2011}  \at{Experimental detection of deterministic turbulence}.
  \jt{Journal of Turbulence}  \bvol{12},  \pg{N23},  \arxiv{arXiv:
  https://doi.org/10.1080/14685248.2011.573792}.
  
\bibitem[Chang {\em et~al.\/}(1993)Chang, Malik, Erlebacher \& Hussaini]{Chang1993}
{\sc \au{Chang, C.-L.}, \au{Malik, M.~R.}, \au{Erlebacher, G.} \& \au{Hussaini, M.~Y.}} \yr{1993}  \at{Linear and nonlinear {PSE} for compressible boundary layers}.
 \org{Technical Report NASA-CR-191537}.
 
\bibitem[Corke \& Mangano(1989)]{Corke1989}
{\sc \au{Corke, T.~C.} \& \au{Mangano, R.~A.}} \yr{1989}  \at{Resonant growth
  of three-dimensional modes in trnsitioning {Blasius} boundary layers}.
  \jt{Journal of Fluid Mechanics}  \bvol{209},  \pg{93--150}.

\bibitem[Craik(1971)]{Craik1971}
{\sc \au{Craik, A. D.~D.}} \yr{1971}  \at{Non-linear resonant instability in
  boundary layers}.  \jt{Journal of Fluid Mechanics}  \bvol{50}~(2),
  \pg{393--413}.

\bibitem[Durbin \& Wu(2007)]{Durbin2007}
{\sc \au{Durbin, P.} \& \au{Wu, X.}} \yr{2007}  \at{Transition beneath vortical
  disturbances}.  \jt{Annual Review of Fluid Mechanics}  \bvol{39}~(1),
  \pg{107--128},  \arxiv{arXiv:
  https://doi.org/10.1146/annurev.fluid.39.050905.110135}.

\bibitem[El-Hady(1988)]{El-Hady1988}
{\sc \au{El-Hady, N.~M.}} \yr{1988}  \bt{Secondary subharmonic instability of
  boundary layers with pressure gradient and suction}.  \org{Technical Report
  NASA-CR-4112}.

\bibitem[Gao {\em et~al.\/}(2011)Gao, Park \& Park]{Gao2011}
{\sc \au{Gao, B.}, \au{Park, D.} \& \au{Park, S.~O.}} \yr{2011}
   \at{Stability analysis of a boundary layer over a hump using parabolized stability equations.}
  \jt{Fluid Dynamics Research}  \bvol{43}~(5),  \pg{055503}.

\bibitem[Herbert(1984)]{Herbert1984}
{\sc \au{Herbert, T.}} \yr{1984} Analysis of the subharmonic route to
  transition in boundary layers.  \bt{In {\em 22nd Aerospace Sciences
  Meeting\/}}. Reno, NV,  \arxiv{arXiv:
  https://arc.aiaa.org/doi/pdf/10.2514/6.1984-9}.

\bibitem[Herbert(1988)]{Herbert1988}
{\sc \au{Herbert, T.}} \yr{1988}  \at{Secondary instability of boundary
  layers}.  \jt{Annual Review of Fluid Mechanics}  \bvol{20}~(1),
  \pg{487--526},  \arxiv{arXiv:
  https://doi.org/10.1146/annurev.fl.20.010188.002415}.

\bibitem[Herbert {\em et~al.\/}(1987)Herbert, Bertolotti \&
  Santos]{Herbert1987}
{\sc \au{Herbert, T.}, \au{Bertolotti, F.~P.} \& \au{Santos, G.~R.}} \yr{1987}
  Floquet analysis of secondary instability in shear flows.  \bt{In {\em
  Stability of Time Dependent and Spatially Varying Flows\/} (ed. \ed{D.~L.
  Dwoyer \& M.~Y. Hussaini})},  \pg{pp. 43--57}.  \publ{New York, NY: Springer
  New York}.

\bibitem[Higham \& Higham(2016)]{Higham2016}
{\sc \au{Higham, D.~J.} \& \au{Higham, N.~J.}} \yr{2016} {\em {MATLAB}
  Guide\/}, 3rd edn.  \publ{SIAM},  \arxiv{arXiv:
  https://epubs.siam.org/doi/pdf/10.1137/1.9781611974669.ch9}.

\bibitem[Issa(1986)]{Issa1986}
{\sc \au{Issa, R.~I.}} \yr{1986}  \at{Solution of the implicitly discretised
  fluid flow equations by operator-splitting}.  \jt{Journal of computational
  physics}  \bvol{62}~(1),  \pg{40--65}.

\bibitem[Jee {\em et~al.\/}(2018)Jee, Joo \& Lin]{Jee2018}
{\sc \au{Jee, S.}, \au{Joo, J.} \& \au{Lin, R.-S.}} \yr{2018}  \at{Toward
  cost-effective boundary layer transition computations with large-eddy
  simulation}.  \jt{Journal of Fluids Engineering}  \bvol{140}~(11),
  \pg{111201--111201--12}.

\bibitem[Joslin {\em et~al.\/}(1993)Joslin, Streett \& Chang]{Joslin1993}
{\sc \au{Joslin, R.~D.}, \au{Streett, C.~L.} \& \au{Chang, C.-L.}} \yr{1993}
  \at{Spatial direct numerical simulation of boundary-layer transition
  mechanisms: {V}alidation of {PSE} theory}.  \jt{Theoretical and Computational
  Fluid Dynamics}  \bvol{4}~(6),  \pg{271--288}.

\bibitem[Kachanov(1994)]{Kachanov1994}
{\sc \au{Kachanov, Y.~S.}} \yr{1994}  \at{Physical mechanisms of
  laminar-boundary-layer transition}.  \jt{Annual Review of Fluid Mechanics}
  \bvol{26}~(1),  \pg{411--482},  \arxiv{arXiv:
  https://doi.org/10.1146/annurev.fl.26.010194.002211}.

\bibitem[Kachanov \& Levchenko(1984)]{Kachanov1984}
{\sc \au{Kachanov, Y.~S.} \& \au{Levchenko, V.~Y.}} \yr{1984}  \at{The resonant
  interaction of disturbances at laminar-turbulent transition in a boundary
  layer}.  \jt{Journal of Fluid Mechanics}  \bvol{138},  \pg{209--247}.

\bibitem[Kim {\em et~al.\/}(2020)Kim, Lim, Kim, Jee \& Park]{Kim2020}
{\sc \au{Kim, M.}, \au{Lim, J.}, \au{Kim, S.}, \au{Jee, S.} \& \au{Park, D.}}
  \yr{2020}  \at{Assessment of the wall-adapting local eddy-viscosity model in
  transitional boundary layer}.  \jt{Computer Methods in Applied Mechanics and
  Engineering}  \bvol{371},  \pg{113287}.

\bibitem[Kim {\em et~al.\/}(2019)Kim, Lim, Kim, Jee, Park \& Park]{Kim2019}
{\sc \au{Kim, M.}, \au{Lim, J.}, \au{Kim, S.}, \au{Jee, S.}, \au{Park, J.} \&
  \au{Park, D.}} \yr{2019}  \at{Large-eddy simulation with parabolized
  stability equations for turbulent transition using {O}pen{FOAM}}.
  \jt{Computers \& Fluids}  \bvol{189},  \pg{108--117}.

\bibitem[Kim(2020)]{Kim2020a}
{\sc \au{Kim, S.}} \yr{2020} Effect of phase difference between instability
  modes on boundary layer transition. Master's thesis, Gwangju Institute of
  Science and Technology, Gwangju, Korea.

\bibitem[Klebanoff {\em et~al.\/}(1962)Klebanoff, Tidstrom \&
  Sargent]{Klebanoff1962}
{\sc \au{Klebanoff, P.~S.}, \au{Tidstrom, K.~D.} \& \au{Sargent, L.~M.}}
  \yr{1962}  \at{The three-dimensional nature of boundary-layer instability}.
  \jt{Journal of Fluid Mechanics}  \bvol{12}~(1),  \pg{1--34}.

\bibitem[Kovacic {\em et~al.\/}(2018)Kovacic, Rand \& Sah]{Kovacic2018}
{\sc \au{Kovacic, I.}, \au{Rand, R.} \& \au{Sah, S.~M.}} \yr{2018}
  \at{Mathieu's equation and its generalizations: {O}verview of stability
  charts and their features}.  \jt{Applied Mechanics Reviews}  \bvol{70}~(2),
  020802,  \arxiv{arXiv:
  https://asmedigitalcollection.asme.org/appliedmechanicsreviews/article-pdf/70/2/020802/6075492/amr\_070\_02\_020802.pdf}.

\bibitem[Lim {\em et~al.\/}(2021)Lim, Kim, Kim, Jee \& Park]{Lim2021}
{\sc \au{Lim, J.}, \au{Kim, M.}, \au{Kim, S.}, \au{Jee, S.} \& \au{Park, D.}}
  \yr{2021}  \at{Cost-effective and high-fidelity method for turbulent
  transition in compressible boundary layer}.  \jt{Aerospace Science and
  Technology}  \bvol{108},  \pg{106367}.

\bibitem[Morkovin(1969)]{Morkovin1969}
{\sc \au{Morkovin, M.~V.}} \yr{1969} On the many faces of transition.  \bt{In
  {\em Viscous Drag Reduction\/} (ed. \ed{C.~Sinclair Wells})},  \pg{pp.
  1--31}.  \publ{Boston, MA: Springer US}.

\bibitem[Nayfeh \& Masad(1990)]{Nayfeh1990}
{\sc \au{Nayfeh, A.~H.} \& \au{Masad, J.~A.}} \yr{1990}  \at{Recent advances in
  secondary instabilities in boundary layers}.  \jt{Computing Systems in
  Engineering}  \bvol{1}~(2),  \pg{401 -- 414}.

\bibitem[Nicoud \& Ducros(1999)]{Nicoud1999}
{\sc \au{Nicoud, F.} \& \au{Ducros, F.}} \yr{1999}  \at{Subgrid-scale stress
  modelling based on the square of the velocity gradient tensor}.  \jt{Flow,
  Turbulence and Combustion}  \bvol{62}~(3),  \pg{183--200}.

\bibitem[Park {\em et~al.\/}(2019)Park, Park, Kim, Kim, Lim \& Jee]{Park2019}
{\sc \au{Park, D.}, \au{Park, J.}, \au{Kim, M.}, \au{Kim, S.}, \au{Lim, J.} \&
  \au{Jee, S.}} \yr{2019} Effect of initial phase difference on subharmonic
  breakdown of boundary layers.  \bt{In {\em AIAA Aviation 2019 Forum\/}},
  \pg{p. 2969}. Dellas, TX,  \arxiv{arXiv:
  https://arc.aiaa.org/doi/pdf/10.2514/6.2019-2969}.

\bibitem[Park \& Park(2013)]{Park2013}
{\sc \au{Park, D.} \& \au{Park, S.~O.}} \yr{2013}  \at{Linear and non-linear
  stability analysis of incompressible boundary layer over a two-dimensional
  hump}.  \jt{Computers \& Fluids}  \bvol{73},  \pg{80 -- 96}.

\bibitem[Park \& Park(2016)]{Park2016}
{\sc \au{Park, D.} \& \au{Park, S.~O.}} \yr{2016}  \at{Study of effect of a
  smooth hump on hypersonic boundary layer instability}.  \jt{Theoretical and
  Computational Fluid Dynamics}  \bvol{30}~(6),  \pg{543--563}.

\bibitem[Saric {\em et~al.\/}(2003)Saric, Reed \& White]{Saric2003}
{\sc \au{Saric, W.~S.}, \au{Reed, H.~L.} \& \au{White, E.~B.}} \yr{2003}
  \at{Stability and transition of three-dimensional boundary layers}.
  \jt{Annual Review of Fluid Mechanics}  \bvol{35}~(1),  \pg{413--440},
  \arxiv{arXiv: https://doi.org/10.1146/annurev.fluid.35.101101.161045}.

\bibitem[Sayadi {\em et~al.\/}(2013)Sayadi, Hamman \& Moin]{Sayadi2013}
{\sc \au{Sayadi, T.}, \au{Hamman, C.~W.} \& \au{Moin, P.}} \yr{2013}
  \at{Direct numerical simulation of complete {H}-type and {K}-type transitions
  with implications for the dynamics of turbulent boundary layers}.
  \jt{Journal of Fluid Mechanics}  \bvol{724},  \pg{480--509}.

\bibitem[Schmid(2007)]{Schmid2007}
{\sc \au{Schmid, P.~J.}} \yr{2007}  \at{Nonmodal stability theory}.  \jt{Annual
  Review of Fluid Mechanics}  \bvol{39}~(1),  \pg{129--162},  \arxiv{arXiv:
  https://doi.org/10.1146/annurev.fluid.38.050304.092139}.

\bibitem[Schmid \& Henningson(2001)]{Schmid2001}
{\sc \au{Schmid, P.~J.} \& \au{Henningson, D.~S.}} \yr{2001} {\em Stability and
  transition in shear flows\/}.  \publ{Springer-Verlag New York}.

\bibitem[Wu(2019)]{Wu2019}
{\sc \au{Wu, X.}} \yr{2019}  \at{Nonlinear theories for shear flow
  instabilities: {P}hysical insights and practical implications}.  \jt{Annual
  Review of Fluid Mechanics}  \bvol{51}~(1),  \pg{451--485},  \arxiv{arXiv:
  https://doi.org/10.1146/annurev-fluid-122316-045252}.

\bibitem[W{\"u}rz {\em et~al.\/}(2012{\natexlab{{\em a\/}}})W{\"u}rz,
  Sartorius, Kloker, Borodulin, Kachanov \& Smorodsky]{Wuerz2012}
{\sc \au{W{\"u}rz, W.}, \au{Sartorius, D.}, \au{Kloker, M.}, \au{Borodulin,
  V.~I.}, \au{Kachanov, Y.~S.} \& \au{Smorodsky, B.~V.}}
  \yr{2012{\natexlab{{\em a\/}}}}  \at{Nonlinear instabilities of a
  non-self-similar boundary layer on an airfoil: {E}xperiments, {DNS}, and
  theory}.  \jt{European Journal of Mechanics - B/Fluids}  \bvol{31},  \pg{102
  -- 128}.

\bibitem[W{\"u}rz {\em et~al.\/}(2012{\natexlab{{\em b\/}}})W{\"u}rz,
  Sartorius, Kloker, Borodulin, Kachanov \& Smorodsky]{Wuerz2012a}
{\sc \au{W{\"u}rz, W.}, \au{Sartorius, D.}, \au{Kloker, M.}, \au{Borodulin,
  V.~I.}, \au{Kachanov, Y.~S.} \& \au{Smorodsky, B.~V.}}
  \yr{2012{\natexlab{{\em b\/}}}}  \at{Detuned resonances of
  {T}ollmien-{S}chlichting waves in an airfoil boundary layer: {E}xperiment,
  theory, and direct numerical simulation}.  \jt{Physics of Fluids}
  \bvol{24}~(9),  \pg{094103},  \arxiv{arXiv:
  https://doi.org/10.1063/1.4751246}.

\bibitem[Xu {\em et~al.\/}(2017)Xu, Lombard \& Sherwin]{Xu2017}
{\sc \au{Xu, H.}, \au{Lombard, J.~W.} \& \au{Sherwin, S.~J.}} \yr{2017}
\at{Influence of localised smooth steps on the instability of a boundary layer}.
\jt{Journal of Fluid Mechanics}  \bvol{817},  \pg{138--170}.

\end{thebibliography}
\end{document}